\documentclass[11pt]{article}
\usepackage{epsfig}
\begin{document}
\setlength{\unitlength}{1mm}
\def\bef{\begin{figure}}
\def\eef{\end{figure}}
\newcommand{\ans}{ansatz }
\newcommand{\be}[1]{\begin{equation}\label{#1}}
\newcommand{\beq}{\begin{equation}}
\newcommand{\ee}{\end{equation}}
\newcommand{\beqn}[1]{\begin{eqnarray}\label{#1}}
\newcommand{\eeqn}{\end{eqnarray}}
\newcommand{\bd}{\begin{displaymath}}
\newcommand{\ed}{\end{displaymath}}
\newcommand{\mat}[4]{\left(\begin{array}{cc}{#1}&{#2}\\{#3}&{#4}
\end{array}\right)}
\newcommand{\matr}[9]{\left(\begin{array}{ccc}{#1}&{#2}&{#3}\\
{#4}&{#5}&{#6}\\{#7}&{#8}&{#9}\end{array}\right)}
\newcommand{\matrr}[6]{\left(\begin{array}{cc}{#1}&{#2}\\
{#3}&{#4}\\{#5}&{#6}\end{array}\right)}
\def\lsim{\raise0.3ex\hbox{$\;<$\kern-0.75em\raise-1.1ex
\hbox{$\sim\;$}}}
\def\gsim{\raise0.3ex\hbox{$\;>$\kern-0.75em\raise-1.1ex
\hbox{$\sim\;$}}}
\def\abs#1{\left| #1\right|}
\def\simlt{\mathrel{\lower2.5pt\vbox{\lineskip=0pt\baselineskip=0pt
           \hbox{$<$}\hbox{$\sim$}}}}
\def\simgt{\mathrel{\lower2.5pt\vbox{\lineskip=0pt\baselineskip=0pt
           \hbox{$>$}\hbox{$\sim$}}}}
\def\unity{{\hbox{1\kern-.8mm l}}}
\def\epr{E^\prime}
\newcommand{\al}{\alpha}
\def\16p{16\pi^2}
\newcommand{\eps}{\varepsilon}
\newcommand{\epsr}{\varepsilon_{ R}}
\newcommand{\epsl}{\varepsilon_{ L}}
\newcommand{\epsrs}{\varepsilon_{s R}}
\newcommand{\epsls}{\varepsilon_{s L}}
\def\ep{\epsilon}
\def\ga{\gamma}
\def\Ga{\Gamma}
\def\om{\omega}
\def\OM{\Omega}
\def\la{\lambda}
\def\La{\Lambda}
\def\al{\alpha}
\newcommand{\ov}{\overline}
\renewcommand{\to}{\rightarrow}
\renewcommand{\vec}[1]{\mbox{\boldmath$#1$}}
\def\tm{{\widetilde{m}}}
\def\mcirc{{\stackrel{o}{m}}}
\def\dem{\delta m^2} 
\def\sint{\sin^2 2\theta} 
\def\tant{\tan 2\theta} 
\def\tanL{\tan 2\theta^L}
\def\tanR{\tan 2\theta^R}
\newcommand{\tanb}{\tan\beta}
\def\dfrac#1#2{{\displaystyle\frac{#1}{#2}}}

%

\begin{titlepage}

\begin{flushright}
DFAQ-01/TH/09 \\ 
DFPD-01/TH/47 \\ 

\end{flushright}

\vspace{2.0cm}

\begin{center}

{\Large \bf 
Probing 
Non-Standard  Couplings of Neutrinos \\
\vspace{0.4cm}
at the Borexino Detector}

\vspace{0.7cm}

{\large \bf Zurab Berezhiani${}^{a,b,}$\footnote{
E-mail address: berezhiani@fe.infn.it }, 
R. S. Raghavan${}^{c,}$\footnote{
E-mail address: raju@physics.bell-labs.com}
and Anna Rossi${}^{d,}$\footnote{
E-mail address: arossi@pd.infn.it } 
}
\vspace{5mm}

{\it ${}^a$ Dipartimento di Fisica, 
Universit\`a di L'Aquila,  
I-67010 Coppito, AQ, and \\
INFN, Laboratori Nazionali del Gran Sasso, I-67010 Assergi, AQ, Italy}

{\it ${}^b$ The Andronikashvili Institute of Physics, 
Georgian Academy of Sciences, \\ 
380077 Tbilisi, Georgia} 

{\it  ${}^c$ Bell Laboratories, Lucent Technologies,  
Murray Hill, New Jersey 07974.
}

{\it  ${}^d$ Dipartimento di Fisica, 
Universit\`a di Padova and 
INFN Sezione di Padova, \\ I-35131 Padova, Italy. 
}
\end{center}

\vspace{10mm}

\begin{abstract}
\noindent
The present experimental status does not exclude weak-strength 
non-standard interactions of neutrinos with electrons.
These interactions can be revealed in solar neutrino experiments.
Our discussion covers several aspects related to this issue.  
First, we perform an  analysis of the Super Kamiokande and SNO 
data to investigate their sensitivity to  such interactions. 
In particular, we show that the $\nu_e$ oscillation into 
sterile neutrinos can be still allowed if $\nu_e$ has extra 
interactions of the proper strength. 
Second, we suggest that the Borexino detector can provide 
good signatures for these non-standard interactions. 
Indeed, in Borexino the shape of the recoil electron spectrum 
from the $\nu~e \to \nu~e$ scattering  
essentially does not depend on the solar neutrino conversion 
details, since most of the signal comes from the mono-energetic 
$^7$Be neutrinos.  
Hence, the partial conversion of solar $\nu_e$  into a 
a nearly equal mixture of  $\nu_\mu$ and $\nu_\tau$, 
as is indicated by the atmospheric neutrino data, 
offers the chance to test extra interactions 
of $\nu_\tau$, or of $\nu_e$ itself. 
\end{abstract}
\end{titlepage}

\setcounter{footnote}{0}

\section{Introduction}
The present experimental data on solar and atmospheric neutrinos 
provide a compelling evidence that neutrinos are massive and mixed. 
In particular, in the context of three standard neutrino states
$\nu_{e,\mu,\tau}$, 
the following paradigm arises for their mixing and mass pattern:

(i) the $\nu_\mu \to \nu_\tau$ oscillation is the dominant 
mechanism for the atmospheric neutrino anomaly (ANA) \cite{venice}; 
$\nu_\mu$ and $\nu_\tau$ are nearly 
maximally mixed, $\theta_{23} \sim 45^\circ$,  
and $\delta m^2_{23}\equiv \delta m^2_{\rm atm} \sim 
3\times 10^{-3} {\rm eV}^2$; 

(ii) the solar neutrino anomaly (SNA) can be interpreted in 
terms of the conversion $\nu_e\to \nu_a$, 
with $\nu_a$ being $\nu_\mu$ or $\nu_\tau$ or a 
combination of them. This fact is clearly confirmed by the   
combined results of the Super-Kamiokande (SK) \cite{SK} 
and the Sudbury Neutrino Observatory (SNO) \cite{SNO}. 
The concrete oscillation scheme  is less clear 
and different conversion mechanisms can be invoked.
Namely, the $\nu_e$ and $\nu_a$ states can be either 
strongly or tinily mixed, 
$25^\circ \lsim \theta_{12} \lsim 60^\circ$ or  
$\theta_{12} \sim  2^\circ$,  
depending on the specific solution adopted, with a corresponding 
$\delta m^2_{12}=\delta m^2_{\rm sol}$ ranging 
from $10^{-12}$ up to $10^{-4}$ eV$^2$ \cite{SNP,BMW}; 

(iii) 
the combined analysis of the atmospheric and solar neutrino 
data points to a small 13 mixing, 
$\theta_{13} \lsim 0.1$, which is also in agreement 
with the data of the CHOOZ experiment \cite{chooz}. 

In this view, the lepton mixing matrix $V$ connecting 
the neutrino flavour eigenstates $(\nu_e, \nu_\mu, \nu_\tau)$ 
with the mass eigenstates $(\nu_1, \nu_2, \nu_3)$, {\it viz.}  
$\nu_\alpha = V_{\alpha i} \nu_i$, can be presented as:
\be{mix}
V = 
\matr{c_{12}}{s_{12}}{0}
{-s_{12}c_{23}}
{c_{12}c_{23}} {s_{23}} 
{s_{12}s_{23}}
{-c_{12}s_{23}} {c_{23}}~, ~~~~~(\alpha =e,\mu,\tau, ~~i=1,2,3) , 
\ee
where $s_{ij}=\sin\theta_{ij}$, $c_{ij}=\cos\theta_{ij}$,   
and $s_{13} =0$ has been assumed.
Hence, $\nu_e$ is mixed with a combination 
$\nu_a= (c_{23}\nu_\mu - s_{23}\nu_\tau) 
\simeq \frac{1}{\sqrt2}(\nu_\mu - \nu_\tau)$, 
which  means that solar neutrinos are converted into a nearly 
equal mixture of $\nu_\mu$ and $\nu_\tau$. In this way, 
the Sun appears to be a copious  source of both {\it muon} and 
{\it tau} neutrinos. However, the solar neutrino 
detectors are not sensitive to the $\nu_\mu$ and $\nu_\tau$ 
fraction individually, 
since in the framework of the Standard Model (SM)  
the neutral current (NC) interactions of these states 
are indistinguishable.  
In particular, experiments like SK which detect the $\nu_a e$ 
elastic-scattering, should not be sensitive 
to the mixing angle $\theta_{23}$.  

However, the present experimental limits still allow 
 neutrinos to have some extra 
(and not necessarily flavour-universal) interactions with the
electrons or nucleons. 
It is very interesting that recently some hint for `non-standard' 
physics of neutrinos has shown up. Namely, the NuTeV results 
on the determination of electroweak parameters show a discrepancy 
with the SM expectation that suggests that muon neutrinos 
have some non-standard couplings with quarks \cite{nutev}. 
Therefore, the occurrence of extra non-standard (NS) interactions 
of neutrinos with the matter constituents 
is an open issue which in our opinion deserves some attention. 

The present experimental data exclude that $\nu_\mu$ can 
have extra NS couplings, at least in the range of 
sensitivity of the solar neutrino detectors. 
However, $\nu_e$ and $\nu_\tau$ are still "experimentally" 
allowed to have significant NS interactions, with 
strength comparable to the Fermi constant $G_F$ 
(for a recent analysis, see \cite{az3}).  
For example, extra interactions of $\nu_\tau$ 
could differentiate its contribution from that of $\nu_\mu$ 
and thus could allow to `identify' the flavour content of $\nu_a$ 
in solar neutrino experiments 
(or, in other words, to test the large $\nu_\mu-\nu_\tau$ 
mixing angle required by atmospheric neutrinos).  
Their impact on the detection cross section of "solar" $\nu_\tau$'s 
was discussed in ref.  \cite{AZ1} some years ago in the context 
of the long wavelength oscillation solution to the SNA. 
In the context of other solutions, one should  bear  
in mind that  non-standard (flavour-diagonal or flavour-changing) 
interactions of $\nu_e$ or $\nu_\tau$ would also affect the neutrino
oscillations in matter \cite{MSW}. 
Their possible implications for solar neutrinos have been 
discussed long ago \cite{FC,BPW,AZ2}. 
(For a more recent analysis, see for example \cite{analysis}.)

In the present paper we concentrate on the NS interactions 
of neutrinos with electrons and discuss their implications 
for solar neutrino experiments. 
We also discuss  another non-standard 
possibility, namely  the existence of a light sterile neutrino $\nu_s$ 
which is mixed to ordinary neutrinos. 
In this case the solar $\nu_e$ could oscillate into 
some state $\nu^\prime$ which is a 
mixture of $\nu_{\mu,\tau}$ and $\nu_s$. 
In the standard picture, 
the comparison between the charged-current (CC) 
measurements performed by SNO and the SK data 
strongly disfavours SNA solutions  due to the $\nu_e$ conversion 
exclusively into $\nu_s$. However, the conversion 
$\nu_e \to \nu^\prime$
is not excluded even for pretty large mixing angle 
$\theta_{as}$ \cite{SNO,BMW}. 

Here, first we study  the impact of neutrino NS interactions 
on the present solar neutrino phenomenology.  
The neutrino NS interactions  with electrons 
can affect the detection reaction $\nu e \to \nu e$ 
in Super-Kamiokande.  
Therefore it is necessary to explore the features that 
may emerge in this case when comparing the SK 
and SNO data. We show that this leads to significant 
constraints on the NS couplings of both  $\nu_e$ and 
$\nu_\tau$.

Finally, we would like to put forward the possibility  
to reveal  non-standard NC interactions 
of $\nu_e$ or $\nu_\tau$ 
with the electron in solar neutrino detectors 
like Super-Kamiokande and Borexino, which are   
sensitive to the $\nu e$ elastic scattering. 
It is well-known that the measurement of the neutrino energy 
spectrum in solar neutrino experiments may be used  
to discriminate the several SNA solutions  
as it does not depend on the solar-model theory.
In general the deformation of the 
energy spectrum is expected to arise from 
the energy dependence of the neutrino survival 
probability $P(E)$ ($E$ is the neutrino energy). 
The effect of spectral deformation due to NS interactions
would then be superimposed to that induced by the energy 
dependence of $P$.  
The advantage of the Borexino experiment is that its signal 
is mainly sensitive to mono-energetic Beryllium neutrino flux.
This makes easier to detect the deformation 
of the recoil electron energy spectrum:  
the effect induced by $P(E)$ 
on the energy spectrum can be nicely `factorised out',    
and thus any specific spectral deformation  
can only be attributed 
to the $\nu$ non-standard   interactions involved in the 
detection reaction $\nu e\to \nu e$.   
Therefore, the capability of Borexino experiment is unique in that 
it could be very sensitive to the spectral distortions  
induced by $\nu$ NS  interactions. 
This experimental evidence would be complementary to that 
achieved by SK and DONUT experiments which both claim 
to have observed charged current $\nu_\tau$ events 
\cite{venice1,donut}.

The paper is organized as follows. 
In Sect. 2 we present the effective Lagrangian describing 
the neutrino NS interactions with electrons, and discuss 
how they could modify the differential cross section 
of $\nu e$ scattering relevant for solar neutrino experiments. 
In Sect. 3 we  consider the relevance of NS interactions 
in confronting the SNO and Super Kamiokande data. 
We shall consider the case of active conversion $\nu_e \to \nu_a$ 
assuming that $\nu_\tau$ or $\nu_e$ have  NS interactions with
electrons and study the allowed parameter space.  
We also analyse the neutrino 
neutrino into the sterile-active admixture, and in particular, 
show that the  
purely sterile conversion $\nu_e\to \nu_s$ is not excluded 
if $\nu_e$ has  NS interactions with the electrons 
in the range allowed by the present experimental limits. 
In Sec. 4 we discuss the implications 
for the Borexino experiment which is aimed to detect 
$^7$Be neutrinos. 
Finally, in Sec. 5 we summarize our findings.

\section{Non-standard interactions and solar $\nu$ detection} 

In the Standard Model, the neutrino elastic scattering  
$\nu_\al e\to\nu_\al e$ is described at low energies by the following 
four-fermion operator ($\nu_\al= \nu_e,\nu_\mu,\nu_\tau$):
\be{SM}
{\cal L}_{\rm SM} = 
-{2\sqrt2 G_F} (\ov{\nu}_\al \gamma^\mu P_L \nu_\al)
\left[\, g_{R} (\ov{e}  \gamma_\mu P_R e) + 
g_{L}(\ov{e}\gamma_\mu P_L e)\,\right] , 
\ee  
where $P_{L,R}= (1\mp \ga^5)/2$ are the chiral projectors and 
$g_R = \sin^2\theta_W$, $g_L=\sin^2\theta_W \pm \frac12$, 
where the lower sign applies for $\nu_{\mu,\tau}$ 
(from $Z$-boson exchange)
and the upper one for $\nu_e$ (from $Z$ and $W$-boson
exchange). 

We assume on phenomenological grounds that neutrinos have  
also extra weak-strength interactions with the electron described 
by the following four-fermion operator:\footnote{
We focus on flavour diagonal NS interactions, 
though in general we could also include   
flavour-changing interactions.} 
\be{NS}
{\cal L}_{\rm NS} = 
-{2\sqrt2G_F} (\ov{\nu}_\al \gamma^\mu P_L \nu_\al)
\left[\, \eps_{\al R} (\ov{e} \gamma_\mu P_R e) + 
\eps_{\al L} (\ov{e} \gamma_\mu P_L e) \, \right] , 
\ee   
where the dimensionless constants $\eps_{\al L,R}$ 
parameterise the strength of the new interactions with respect to $G_F$. 
These interactions are not necessarily flavour universal, 
and can be different for $\nu_\al=\nu_e,\nu_\mu,\nu_\tau$.   
As for low energy $\nu_\al e\to \nu_\al e$ scatterings   
are concerned, the NS interaction effects entail the following redefinition 
of the coupling constants $g_{R},g_{L}$ in (\ref{SM}): 
\be{NCnutau}
g_{R} \to \tilde{g}_{\al R} = g_{R} + \eps_{\al R} ~, ~~~~~~
g_{L} \to \tilde{g}_{\al L} = g_{L} + \eps_{\al L} ~. 
\ee
Sometimes the four-fermion interaction (\ref{SM}) is parameterised in 
terms of  the vector and axial 
constants $g_{V}= g_L + g_R$, $g_{A}= g_L- g_R$. In a similar way, 
we can define the  NS couplings, 
$\eps_{\al V} = \eps_{\al L} + \eps_{\al R}$ and 
$\eps_{\al A} = \eps_{\al L} - \eps_{\al R}$ and 
consequently $\tilde{g}_{\al V}= \tilde{g}_{\al L} + \tilde{g}_{\al R}$, 
$\tilde{g}_{\al A}= \tilde{g}_{\al L} - \tilde{g}_{\al R}$.

Let us discuss now the experimental bounds 
on the NS interactions (\ref{NS}). 
At present, the strongest limits 
are posed by  $\nu_\mu e$ scattering experiments, 
which constrain 
NS interactions of $\nu_\mu$ with electrons, 
$|\eps_{\mu R,L}| < 0.02$ or so \cite{CHARM2}. 
However, regarding the electron neutrino, 
the existing laboratory bounds from low-energy 
$\nu_e e$ and $\bar{\nu}_e e$ scattering are rather weak 
and allow significant deviations from the SM predictions,  
while for $\nu_\tau$ there are no direct limits 
from  low-energy experiments. 
As was recently discussed in \cite{az3},  
neutrino NS interactions with electrons 
can be constrained by the LEP measurements of the 
$e^+ e^- \to \nu\bar{\nu}\gamma$   cross section.   
However, these limits still allow 
neutrino NS interactions to have   a strength comparable 
to the Fermi constant $G_F$.

Finally,  phenomenological bounds on neutrino NS 
interactions of $\nu_\tau$   
can be also obtained from the atmospheric neutrino data. 
These bounds apply only   
to the NS vector-coupling 
$\eps_{\tau V}=\eps_{\tau L} + \eps_{\tau R}$ 
which would induce 
matter effects in the $\nu_\mu\to \nu_\tau$ oscillation. 
Depending on the method adopted to analyse the data, 
for the flavour-diagonal coupling  with electrons 
the bounds obtained in the literature are 
$\eps_{\tau V} \lsim 0.2$ at 99\% C.L.  \cite{fornengo} and 
$\eps_{\tau V} \lsim 0.5$ at 90\% C.L. \cite{FLM}.
However, all these limits are not valid if $\nu_\tau$ has also 
some NS interactions  with quarks which could cancel out  
the effects of non-zero $\eps_{\tau V}$.  
Thereby, in the following we prefer to keep an open mind 
and consider $\eps_{e L,R}$ and $\eps_{\tau L,R}$ 
to be constrained only by laboratory neutrino experiments.

The existing experimental limits on the parameters 
$\eps_{eR,L}$ and $\eps_{\tau R,L}$ are shown in Fig.~\ref{p3}.
Here we also draw the solid line (denoted by $g_V^{\rm SM}$) 
along which $\eps_{e(\tau) V}=0$, and thus the 
neutrino vector coupling to electrons is the same 
as in the SM. 
Therefore, along this direction of the parameter space 
 long-baseline experiments would not be sensitive to such NS interactions  
since the neutrino potential in matter is the same  as in the SM. 
For the same reason, $\eps_{\tau V}=0$ implies no matter effects 
on the  atmospheric $\nu_\mu \to \nu_\tau$ oscillation 
pattern.  

Due to the SM multiplet structures, 
the NS interactions of neutrinos (\ref{NS})  
in general emerge  with other interactions 
involving their charged partners in the weak isodoublets 
\cite{BGP} (for more 
details and  a more model independent approach see \cite{az3}). 
As a matter of fact, the most severe laboratory bounds  
 rather  apply to the charged lepton NS interactions, and    
their strength can be at most several percents of $G_F$. 
However, as it was shown in ref. \cite{az3},   these limits 
cannot be directly translated into limits for the NS couplings 
of neutrinos, for some  conspiracy among  the  
contributions of different operators cannot be excluded  and 
 neutrino NS couplings  can be  sizeable while charged-lepton NS 
interactions can be properly suppressed.

Now we discuss the role of the extra interactions (\ref{NS}) 
for the detection of solar neutrinos via their 
elastic scattering off electrons, which is relevant for 
Super-Kamiokande and Borexino experiments.
For the sake of completeness, we  consider 
the most general case when the solar $\nu_e$ 
oscillates into the active-sterile combination 
$\nu^\prime = s_{as} \nu_a + c_{as}\nu_{s}$  
($s_{as}= \sin\theta_{as},  c_{as}=\cos\theta_{as}$),  
where the  active component is  
$\nu_a = c_{23}\nu_\mu - s_{23}\nu_\tau$.    
The case $s_{as} =1$ corresponds to the 
$\nu_e\to \nu_a$ (fully active) conversion, whereas 
$s_{as} =0$ corresponds to the 
$\nu_e\to \nu_s$ (fully sterile) conversion.\footnote{
As far as the solar neutrino detection via $\nu e$ scattering,  
the admixture of the sterile neutrino 
with the active state $\nu_a$ can be regarded  
as a formal redefinition 
of the SM couplings both for $\nu_\mu$ and $\nu_\tau$,  i.e 
$g^2_{R,L} \to \tilde{g}^2_{R,L} = s_{as}^2 g_{R,L}^2$, 
and then can mimic the presence of  NS couplings,  
$\eps_{\mu (\tau)R,L}= -(1-s_{as}) g_{R,L}$. 
}
The expected energy spectrum of the recoil electrons is given 
by the following expression: 
\be{spec}
S(T) = \sum_i \phi_i 
\int dE \la_i(E) \left [
\dfrac{d\tilde{\sigma}_{\nu_e}}{dT} P(E) + 
s_{as}^2\left(c^2_{23} \dfrac{d\tilde{\sigma}_{\nu_\mu}}{dT} 
+s^2_{23} \dfrac{d\tilde{\sigma}_{\nu_\tau}}{dT}\right) 
\left[ 1 -  P(E)  \right] \right] , 
\ee 
where $\phi_i$ are the fluxes of the 
solar neutrino sources which can be relevant for the signal 
($i= ^8$B, $^7$Be, $pp$ etc.), 
$\la_i(E)$ are the corresponding energy spectra (normalized to unity), 
and $P(E)$ is the generic survival probability of solar $\nu_e$ 
with the energy $E$.  
The differential cross sections are ($\alpha =e,\mu, \tau$):
\be{diff}
\dfrac{d\tilde{\sigma}_{\nu_\alpha}(E,T)}{dT} = 
\frac{2}{\pi} G^2_F m_e  \int^{{T}'_{\rm max}}_0 dT'~  
\rho(T,T') \left[ \tilde{g}^2_{\al L} + 
\tilde{g}^2_{\al R} \left(1 - \frac{T'}{E}\right)^2 
- \tilde{g}_{\al L}\tilde{g}_{\al R} \frac{m_e T'}{E^2}
\right] ,   
\ee
where $T$ is the  `reconstructed' recoil electron kinetic energy,  
$T'$ is the `true' value given by the kinematics and ranging as 
$0\leq T'\leq {T}'_{\rm max} =\frac{E}{1 +m_e/2E}$, and 
$\rho(T,T')$ is the resolution function (explicitly given below). 
The possible non-standard interactions are included 
in the coupling constants $\tilde{g}_{\al L}$ and 
$\tilde{g}_{\al R}$ according to the re-definition 
shown in eq.~(\ref{NCnutau}). 
In the evaluation of $S(T)$ we also include the  
standard radiative corrections to the SM couplings 
(\ref{SM}) \cite{BKS}.

\section{SNO {\it versus} Super-Kamiokande}

The  SNO and SK experiments are both dominated by the 
$^8$B neutrinos (neglecting the smaller flux of the {\it hep} 
neutrinos). 
The SNO experiments detects solar neutrinos with most significance 
via the CC reaction $ \nu_e d \to p p e^- $ 
which is sensitive only to the $\nu_e$, while the SK 
is sensitive to all neutrino flavours through the 
$\nu_\alpha e\to \nu_\alpha e$ scattering, though 
the contribution of $\nu_{\mu,\tau}$ is about a factor 6 
smaller than that of $\nu_e$.  
The measured fluxes normalized to the SSM prediction,   
$\phi_B^{\rm SSM}=5.05 \times 10^6$ cm$^{-2}$s$^{-1}$ \cite{BP00}, 
are:
\be{B8}
Z_{\rm SK} = 0.459\pm 0.017 , ~~~~~~~ 
Z_{\rm SNO} = 0.347\pm 0.027 , 
\ee 
from where we  infer that the relative gap between the two signals 
may  be filled by  $\nu_{\mu,\tau}$-induced events in SK due to the 
solar neutrino conversion $\nu_e \to \nu_a$. This is the feature which 
disfavours the $\nu_e$ depletion exclusively 
into sterile neutrinos, in which case 
$Z_{\rm SK}= Z_{\rm SNO}$ would be expected.  
 
To discuss the impact of the NS interactions we take the 
following point of view.
First, in view of the absence of a significant spectral 
deformation as reported by both SK and SNO experiments,  
we assume that the $\nu_e$ survival probability $P$  
does not depend on the energy, at least in the range 
explored by SK and SNO ($E_e > 5$ MeV).\footnote{This 
assumption is indeed well 
justified in the context of the averaged vacuum oscillation 
solution or the large mixing angle MSW solution.} 
Second, as the $^8$B flux is determined by the SSM with an accuracy of 
20\% or so, we treat it as a free parameter and parameterise 
it as $\phi_{B} = f_B \phi_B^{\rm SSM}$. 
Then the expected signals can be expressed as follows:
\be{Zss1}
%
Z_{\rm SK} = f_B\left[r_e P+
R^{\rm SM}_{\mu/e} 
s_{as}^2 \left(c^2_{23}r_\mu + s^2_{23} r_\tau \right)
(1-P) \right] , ~~~~~~Z_{\rm SNO} = f_B P ,  
\ee
where the `averaged' cross section $R_{\al}$:
\be{Rx}
R_\alpha 
=\int {\mbox d} E \la_B(E) \tilde{\sigma}_{\nu_\alpha}(E) , ~~~~~~~  
 \tilde{\sigma}_{\nu_\alpha} = 
\int^{T_{\rm max}}_{T_{\rm th}} dT~ 
\dfrac{ d\tilde{\sigma}_{\nu_\alpha}(E, T)}{ d T}
\ee
are understood to include also the contributions from neutrino NS
interactions, and the resolution function accounted in
$\tilde{\sigma}_{\nu_\alpha}$ is
\be{rf1}
 \rho(T,T') = \dfrac{1}{ \sqrt{2\pi} \sigma}
\exp \left[-\frac{ (T-T')^2}{2 \sigma^2}\right] ,  ~~~~~~~~
\sigma = \sigma_0 \left( \frac{T'}{\rm MeV} \right)^{1/2} .
\ee
For Super-Kamiokande $\sigma_0=0.47$ MeV, and the 
`reconstructed' kinetic energy $T$ 
ranges  between the threshold value 
$T_{\rm th} = (5~{\rm MeV} -m_e)$ 
and ${T}_{\rm max} = (20~{\rm MeV} -m_e)$. 
In eq.~(\ref{Zss1}) the factors 
$r_\al = R_\al/R_\al^{\rm SM}$  ($\al=e,\mu,\tau$)  
parameterise the cross section deviation from the SM  
expectation $R^{\rm SM}_\al$ 
(i.e. from the limit $\eps_{\al L,R}\to 0$), and 
$R_{\mu/e}^{\rm SM} = R^{\rm SM}_\mu/R^{\rm SM}_e$.\footnote{
As it was already mentioned, due to the severe experimental 
limits on $\eps_{\mu L,R}$ \cite{BPW}, we assume that 
$\nu_\mu$ have only SM interactions and thus we take $r_\mu=1$.}
In particular, for the electron energy threshold $E_e =5$ MeV
it is $R^{\rm SM}_{\mu/e} \approx 0.158$.

For  given values of the parameters $r_e$ and 
$r^\prime=s_{as}^2 (c_{23}^2 r_\mu + s_{23}^2 r_\tau)$, 
the combination of the SK and SNO signals determines both 
the $\nu_e$ survival probability and the Boron neutrino flux: 
\be{fB-P} 
P^{-1} = 1 + \frac{1}{r^\prime R^{\rm SM}_{\mu/e}}
\left(\frac{Z_{\rm SK}}{Z_{\rm SNO}} -r_e\right) , ~~~~~~
f_B = P^{-1} Z_{\rm SNO} . 
\ee  
In particular, in the system of three standard neutrinos, 
with $s_{as}=1$ and $r_{e,\mu,\tau} =1$,
we get $P= 0.33 \pm 0.10$ and $f_B = 1.06 \pm 0.42$, 
in  agreement with the SSM prediction   
$f_B=0.84 - 1.2$.  

Below we analyse both the case of $\nu_e$ conversion into 
the pure active state $\nu_a$ ($s_{as}=1$), and the case of 
$\nu_e$ conversion into  
the admixture of $\nu_a$ and $\nu_s$ ($s_{as}<1$). 
For simplicity, we assume that either $\nu_e$ or 
$\nu_\tau$ have  NS interactions, though in 
general  $\nu_e$ and $\nu_\tau$ may both have NS interactions.

\subsection{Without sterile neutrino $(s_{as}=1)$ }

$\bullet$ 
Consider first the case when only $\nu_e$ has 
extra NS interactions:  
$\eps_{e R,L}\neq 0$, $\eps_{\tau R,L} = 0$.  
Then the SK signal is insensitive 
to the flavour content of $\nu_a$ (i.e. to 23 mixing angle)
and so we have  
$Z_{\rm SK}=f_B [r_e P+ R^{\rm SM}_{\mu/e}(1-P)]$. 
Since the standard $\nu_a$ 
contribution to the SK signal is small, 
$R^{\rm SM}_{\mu/e}\approx 0.16$, 
we expect that too a  big deviation from $r_e=1$, i.e. 
a  $\nu_e$-induced contribution to   
 the SK signal too far 
from the standard expectation, would cause 
an unacceptable  mismatch between the SNO and SK signals. 
The results are shown in Fig.~\ref{p2} (left panel).
By comparing the $1\sigma$ allowed range of the SNO and SK 
signals, we see that within the SSM uncertainties for $f_B$ 
(delimited by dashed vertical lines), we obtain 
rather strong upper and lower bounds 
on $r_e$:  $0.85 \leq r_e \leq 1.25$. 
The allowed range for $r_e$ can be translated 
into allowed space for the parameters $\eps_{e R,L}$. This is visualized  
in Fig.~\ref{p3} (left panel). The parameter space 
allowed by the SK/SNO analysis is that delimited by 
the elliptical contours $r_e=0.85$ and $r_e=1.25$.  

On the same Fig.~\ref{p3} we also show  the LSND limits 
obtained from the $\nu_e e$ elastic scattering (shaded area) 
and the LEP limits derived from 
$e^+e^-\to \nu\bar{\nu} \gamma$ process 
(area between dotted circles). 
We observe that the limits from the solar neutrino 
experiments are complementary to those from laboratory experiments,   
and in combination with the latter, provide very strong 
constraints on the extra interactions of $\nu_e$. 
Namely, there are two `disconnected' allowed regions. 
The upper one, an horizontally elongated strip 
localized around the SM point ($\eps_{eR}, \eps_{eL}$)=(0,0) delimited as:
\beqn{sna-e}
& -0.06 \leq \eps_{eL} \leq 0.15 , ~~~~~~~~&({\rm any}~ \eps_{eR}),
\nonumber \\
& -0.5 \leq \eps_{eR} \leq 0.8 , ~~~~~~~~& ({\rm any}~ \eps_{eL}) .
\eeqn 
The range for $\eps_{eR}$ is in fact reduced to 
$-0.5 \leq \eps_{eR} \leq 0.6$ (for any $\eps_{eL}$) 
if  also the reactor $\bar{\nu}_e$ 
bounds are considered \cite{az3}. In the following we 
shall most focus on the SM neighborhood $(\eps_{eR}, \eps_{eL})=(0,0)$ 
and so consider as  allowed region\footnote{
In Fig.~\ref{p3} we superimpose the parameter space allowed 
by LSND and LEP data at 99\% C.L. with that allowed by SK/SNO data 
at 68\% C.L. Though the corresponding comparison may look unsatisfactory, 
notice that it may be acceptable for the bounds inferred on $\eps_{e R, L}$, 
especially around the SM point, we are mostly interested in. 
On the other hand, we are aware that 
by taking the SK/SNO 99\% C.L. contours for   $\eps_{\tau R, L}$ 
would yield looser lower bound on $\eps_{\tau  L}$   
than that reported below in eq.~(\ref{sna-tau}). However, 
we are not interested in the negative range of  $\eps_{\tau  L}$.
} 
that parameterised by
$\eps_{eL}=0, |\eps_{eR}|\leq 0.5$.  
The lower region is the smaller strip elongated 
along the direction $\eps_{e L} \sim -1.4$, 
with minor statistical significance.  
In addition, it is not appropriate for the MSW solution 
since strongly diminishes (or even makes negative) 
the matter potential of $\nu_e$. Therefore, in the 
following analysis we disregard this parameter region. 

$\bullet$ 
The same  procedure can be applied when 
only $\nu_\tau$ has  NS interactions: 
$\eps_{\tau R,L}\neq 0$, $\eps_{e R,L}=0$. 
In this case we have 
$Z_{\rm SK}= f_B [P + r^\prime R^{\rm SM}_{\mu/e}(1-P)]$, 
where the $\nu_\tau$ contribution depends 
on the 2-3 mixing angle of neutrinos,  
$r^\prime = c_{23}^2 + s_{23}^2 r_\tau$. 
For the sake of definiteness, we consider the case of 
maximal 2-3 mixing, as motivated by 
the atmospheric neutrino observations, which results in 
$\nu_a = (\nu_\mu -\nu_\tau)/\sqrt{2}$, and thus  
$r^\prime = \frac12(1+r_\tau)$. 
The results are shown in Fig.~\ref{p2} (right panel).
We observe that for large enough Boron flux, $f_B \gsim 1.2$,  
$r_\tau$ could go to zero as the SK signal would be recovered 
by the standard contribution of $\nu_\mu$. 
Therefore, in this case   
$\nu_\tau$ is allowed to behave  like a `sterile' state 
($r_\tau=0$ means that the  
standard couplings to electrons are exactly canceled 
by the NS couplings, $\eps_{\tau R,L} = -g_{R,L}$).  
On the contrary, for $f_B < 1.3$ the $\nu_\tau$ cross section 
could be larger than the SM one.  
We observe that within the SK/SNO signal and the SSM 
uncertainties, 
$r_\tau$ could vary in the range from 0 to about 3. 
Correspondingly, the allowed range\footnote{
This bounds on $r^\prime$ can be directly translated 
into  bounds on $r_\tau$ for arbitrary $\theta_{23}$ and 
$\theta_{as}$ mixing angles. 
In particular, in case of the exclusively $\nu_e\to \nu_\tau$ 
conversion, $r^\prime = r_\tau$. 
}  
of $r^\prime$ is 
$0.5 \lsim r^\prime \lsim 2$. 

The contours of $r_\tau$ as a function of $\eps_{\tau R,L}$, 
are shown in Fig.~\ref{p3} (left panel).  
The comparison with  the LEP limits shown on the same Fig.~\ref{p3} 
(shaded region) demonstrates  that the $1\sigma$-limit $r_\tau\lsim 3$ 
inferred from the SK/SNO signal analysis, considerably 
restricts the allowed parameter space of $\eps_{\tau R,L}$. 
Most conservatively, ignoring any correlation, the allowed range for 
individual parameters is  
\beqn{sna-tau}
& -0.23 \leq \eps_{\tau L} \leq 0.45  ~~~~~ &({\rm any}~\eps_{\tau R} ), 
\nonumber \\
&-0.45 \leq \eps_{\tau R} \leq 0.65 , ~~~~~ &({\rm any}~ \eps_{\tau L}) .  
\eeqn 
In conclusion, even if  the bounds on $\nu_\tau$ NS  interactions 
obtained by  solar neutrino experiments are comprehensively 
much weaker than the LEP bounds, they are complementary  
and, in particular, cut out the parameter 
space corresponding to large negative values of $\eps_{\tau L}$.

\subsection{The active-sterile conversion $(s_{as} < 1)$  }

$\bullet$ 
Let us first analyse the conversion 
$\nu_e \to \nu^\prime = s_{as} \nu_a + c_{as}\nu_{s}$, 
where  $\nu_e$ and $\nu_\tau$ are assumed to 
have only SM interactions. Hence  in eq.~(\ref{Zss1}) 
the SK signal gets simplified into
$Z_{\rm SK} = f_B [P + s_{as}^2 (1-P) R^{\rm SM}_{\mu/e} ]$.
So upon comparing the SNO and SK signals, for a given $f_B$ 
we can constrain the magnitude of the  active fraction, $s_{as}^2$.   
In Fig.~\ref{p1} we show the allowed 1$\sigma$  
region in the plane ($f_B, s^2_{as}$) (left panel).   
As we could expect, $s_{as}^2$ cannot vanish even in 
the asymptotic regime of very large $f_B$,  and so 
the pure sterile conversion is strongly disfavoured. 
For the `central value' of the Boron flux ($f_B=1$) 
we find $s^2_{as} \geq 0.7$, while for $f_B = 1.2$
also $s^2_{as} = 0.5$ can be allowed which would 
correspond to maximal $\nu_a -\nu_s$ mixing.\footnote{Our 
allowed region somewhat differs from other analyses as 
that in \cite{BMW} because of different `fitting' procedure.}
As for the $\nu_e$ survival probability, $P=Z_{\rm SNO}/f_B$, 
the SSM range $f_B=0.84 - 1.2$ 
implies that $0.32\lsim P \lsim 0.38$ (at 1-$\sigma$).

$\bullet$ 
Let us now assume  that $\nu_e$ have some NS couplings 
with the electron, i.e. $\eps_{e R,L} \neq 0$. 
Now we have to refer to the more general expression 
for $Z_{\rm SK}$ displayed in eq.~(\ref{Zss1}), 
with $r^\prime = s_{as}^2$. Then for given $f_B$, 
we have $P=Z_{\rm SNO}/f_B$, and so the first  
equation in eq.~({\ref{fB-P})) describes the correlation 
between $r_e$ and $r^\prime=s_{as}^2$.
The corresponding isocontours are presented in Fig.~\ref{p1} 
(right panel).\footnote{Notice, that in fact the same 
correlation can be used for understanding the case when 
both $\nu_e$ and $\nu_\tau$ have NS coolings, e.g. 
when $r_e\neq 1$ and $r^\prime = (1+r_\tau)/2\neq 1$. 
}    
Now we observe that 
by increasing the $\nu_e$ cross section, 
in the range $r_e \sim 1.2 - 1.5$, the completely sterile 
conversion ($s_{as} =0$) can be also allowed. 
Notice that for $s^2_{as} \to 0$ all curves in Fig.~\ref{p1},
corresponding to different values of $f_B$, 
are attracted to the `fixed' points, $r_e=1.2, 1.5$.  
The reason is that for $r^\prime=0$, the two eqs. (\ref{Zss1}) 
degrade to $Z_{\rm SNO}=f_B P$ and  
$Z_{\rm SK}=f_B P r_e$, 
which can be satisfied, independently of $f_B$, 
if and only if $r_e = Z_{\rm SK}/Z_{\rm SNO}$. 
For the central values of $Z_{\rm SK}$ and $Z_{\rm SNO}$ 
this would correspond to $r_e=1.32$, 
while within $1\sigma$ uncertainty in (\ref{B8}), we can have 
$r_e=1.2-1.5$. 
As we see from the Fig. \ref{p3}, this possibility is 
still allowed by the LSND data on $\nu_e e$ scattering 
and the collider data on $e^+ e^- \to \nu_e\bar{\nu}_e\gamma$ 
cross section. 

The same exercise has been repeated by allowing  $\nu_\tau$ 
to have the NS interactions. However, in this case we do  
not get more information than what we obtained in Sec. 3.1. 
The admixture with the sterile 
neutrino  simply re-scales the allowed range for 
the parameter $r^\prime =(1+r_\tau)/2$, {\it viz.} $0.5\lsim r^\prime 
\lsim 2$, just by the factor $s_{as}^{-2}$.

\subsection{Spectral deformation at Super-Kamiokande} 

We have seen that the solar neutrino experiments contribute  further in 
constraining the parameter
$\eps_{eL}$, with respect to the LSND experiment.
For example,  for $\eps_{eR}=0$, the values $r_e=0.85,1,1.25,1.5$ 
correspond to $\eps_{eL} = -0.05, 0, 0.07, 0.16$, respectively,  
to be compared with the LSND limits 
$-0.04 < \eps^e_L < 0.08$ ($68\%$ C.L.) and 
$-0.15 < \eps^e_L < 0.17$ ($99\%$ C.L.). 
On the other hand, the SK/SNO analysis provides much looser 
limits on $\eps_{eR}$. Therefore, as already discussed, we consider 
the range $|\eps_{eR}|\leq 0.5$ obtained by comparing 
the LSND and LEP data with the data from $\bar{\nu}_e e$ scattering at 
reactors \cite{az3}.  
The different sensitivity of the solar neutrino experiments 
to $\eps_{eL}$ and $\eps_{e R}$ 
(or analogously to $\eps_{\tau L}$ and $\eps_{\tau R}$)  is  
easy to understand. 
The magnitude of the total signal in SK, as well as of the cross section 
measured at  LSND,
is mostly controlled by the $\tilde{g}^2_{eL}$-term in the 
cross section. As a matter of fact 
 both the LSND data and the SK/SNO analysis imply, around the 
point $(\eps_{eR},\eps_{eL}) =(0,0)$, that $\tilde{g}_{eL}$ should
not deviate too much  
from the SM prediction
$g_L = \sin^2\theta_W + 1/2$. On the other hand, 
as one can see from eq.~(\ref{diff}), 
the parameter $\tilde{g}_{eR}$ controls the energy dependence 
of the recoil electron spectrum.
Therefore, at this point, we may wonder about the implications of NS 
interactions with electrons for the energy spectrum measured at SK 
(since in the above we have considered only the global rates).  
Without entering into a detailed $\chi^2$ analysis, 
which is beyond our scope, 
we have performed a scanning of the 
 $\eps_{eR}, \eps_{eL}$ and $P$ space, and 
we have realized that 
$\eps_{eR}$ is poorly constrained. 
For the sake of demonstration,  
in Fig.~\ref{p4} (left panel) we show the spectral 
deformation expected at  SK for several values of $\eps_{eR}$ in the interval 
$-0.4 \div 0.4$, taking $\eps_{eL} =0$. We can   
observe quite small deviations from the SM case.
On the right panel of the same Fig.~\ref{p4}
we show the spectral deformations for different 
values of $\eps_{\tau R}$ within the experimentally 
allowed interval.  
The reason of 
this poor sensitivity of the SK recoil electron 
spectrum  to $\eps_{e(\tau) R}$ is that it is smoothed out 
by the integration over the {\it continuous} 
Boron neutrino spectrum. 
We will see in next section that the {\it mono-energetic} 
character  of the Beryllium neutrinos makes the Borexino 
detector more sensitive to these NS couplings.

\section{Predictions for Borexino }

We consider as prototype for our discussion the Borexino experiment 
which is aimed to detect mono-energetic  
$^7$Be neutrinos via $\nu e$ elastic scattering \cite{BX}. 
Eq.~(\ref{spec}) shows the advantage of using mono-energetic neutrinos, 
with $\lambda(E) = \delta(E-E_0)$ and $E_0$ is the neutrino line energy.  
In this case the smearing of the electron energy distribution 
due to the integration over  the neutrino spectrum is absent.
Therefore, while the expected global rate substantially 
depends on the $\nu_e$ survival probability $P=P(E_0)$, 
the shape of the recoil electron spectrum 
does not get substantially deformed 
in the SM scenario ($\eps_{\al L,R} =0$).  
Thus, for mono-energetic neutrinos 
a distortion of the electron energy distribution 
would be an unambiguous evidence of neutrino non-standard 
interactions. 
This would represent a unique signature of new physics  
that may be provided by solar neutrino experiments.  

The Beryllium neutrino flux, with $E_0=0.862$ MeV, 
can be measured by exploring the energy window 
$T = 0.25 - 0.66$ MeV for the recoil electron: 
0.25 is the attainable detection threshold and 
0.66 is the Compton edge. In this window, 80\% of the signal 
is from the $^7$Be neutrinos, while the rest comes from the CNO 
and $pep$ neutrinos. 
In fact, both the edges  get  smeared when the energy 
resolution is taken into account and the lower part of the 
spectrum is arisen by the $pp$ and $^7$Be$(0.38 {\rm MeV})$ 
neutrinos. 
These features are clearly apparent in Fig.~\ref{p5} 
(upper left panel) where the energy distributions of the expected 
events are plotted. 
The resolution function used is of the form given in eq.~(\ref{rf1}) 
with width $\sigma_0 = 57.7$ KeV  \cite{taup95}. 
We see that in the interval $T=0.3-0.6$ MeV the spectral 
shape is quite regular with almost constant slope. 

Now let us turn to neutrino oscillations. 
The several  scenarios proposed to explain the SNA predict 
different survival probabilities for the beryllium 
neutrinos.\footnote{In order to avoid confusion, we must 
note that the $^7$Be neutrino survival probability $P$ 
of this section in general should be different from the 
$^8$B neutrino survival probability 
also denoted by $P$ in the previous section.} 
For example the suppression can be quite strong,  
$P\lsim 0.2$ in case of small-mixing angle MSW conversion, 
but can be weaker, $P\sim 0.3 - 0.7$ 
in case of vacuum oscillation or large-mixing angle 
MSW conversion. 
On the other hand, as mentioned,   
neutrino conversions are expected not to distort the energy 
distribution as long as $\nu_{\mu,\tau}$ have only SM interactions. 
Then the energy spectrum will be just re-scaled with a slope 
which will be in between  that of the SSM distribution and 
that expected in case of complete depletion of solar $\nu_e$. 
The energy distributions shown in the upper right panel of Fig.~\ref{p5} 
for different values of $P$, assuming that all neutrinos 
have the standard model  couplings, can help to figure out 
the standard situation.  
Notice that when 
all $\nu_e$'s are  converted into  $\nu_a$ ($P=0$) 
the distribution becomes flatter in the range 
$T=(0.3-0.6)$ MeV (lowest curve).

We now discuss  how this picture gets modified when NS interactions 
of neutrinos are switched on. In the following, in view of the 
results obtained in the previous section,  the parameters 
$\eps_{e, \tau L}$ are set to zero, whereas the parameters 
$\eps_{e, \tau R}$ are varied into the corresponding allowed ranges.
The lower plots  in Fig.~\ref{p5} represent the expected event 
distribution at Borexino as a result of $\nu_e\to 
(\nu_\mu -\nu_\tau)/\sqrt{2}$  transition for different values of $P$ and 
for $\eps_{eR} =0.4, -0.3$  and 
$\eps_{\tau R} =0.6, -0.3$.
By comparing the analogous SM cases (upper plots), 
we observe  for positive $\eps_{e, \tau R}$  a strong deformation 
of the energy distribution, with a noticeable negative slope 
 accompanied by an increase of the number of events. 
We also see that the case in which only  $\eps_{eR}$  
is switched on can be distinguished by the one in which 
only $\eps_{\tau R}$  is  on. In the former case the decreasing 
of the spectrum is even sharper.  
The presence of the background in the lower-energy end may render 
hard to observe the part mostly deformed. However, due to the exponential 
decay of the background\footnote{The shape of the background 
increases from 0.25 up to 0.4 MeV
and for $T\gsim 0.4$ MeV exponentially decay  \cite{meroni}.} 
it is still possible to reveal 
the deformation above (0.4 - 0.45) MeV.
On the other hand, for negative $\eps_{e, \tau R}$ 
the previous strong spectral deformation 
is  lost, but the spectrum becomes flatter  
as $P$ becomes smaller.

We have then made more apparent the comparison 
between the  SM picture and that 
in which neutrinos  have  also NS interactions. 
We have  compared  the electron energy 
spectra obtained for different $\eps_{e, \tau R}$ 
but normalized to the same number of total events $N_{\rm tot}$ 
in the energy window $T= 0.3-0.6$ MeV 
(where essentially only $^7$Be neutrinos contribute). 
For definiteness, we have  $N_{\rm tot}=23$ corresponding to $P=0.5$ 
and $\eps_{e, \tau R}=0$. Then   for each value of $\eps_{e, \tau R}$
the corresponding $\nu_e$ or $\nu_\tau$ cross section is evaluated, i.e 
the parameter $r_{e, \tau}$, while  
$P$ is determined by imposing  $N_{\rm tot}=23$. The energy spectra 
(divided by the SSM prediction, i.e. taking 
 $P=1$, $\eps_{e, \tau R}=0$) 
are displayed in  Fig.~\ref{p6} (upper panels).
For example, for $\eps_{e R} > 0$ (left panel) the spectrum is 
very deformed with a negative slope. On the other hand, 
for $\eps_{e R} \sim -0.3$ the slope becomes positive, 
though the deformation is less pronounced.  

Notice that in the case of the SK spectra, analysed in Fig.~\ref{p4} 
the slope never becomes positive, even for negative values of 
either $\eps_{e R}$ or $\eps_{\tau R}$. 
This different behaviour can be qualitatively  understood by 
considering the derivative of the $\nu e$ differential cross 
section:  
\be{deriv} 
\frac{{ d}}{{ d} T}
\left[\frac{d \sigma_{\nu_\al}(E,T)}{d T}\right] 
\propto -2 \left(1- \frac{T}{E}\right) - 
\frac{\tilde{g}_{\al L}}{\tilde{g}_{\al R}} \frac{m_e}{E} .
\ee
This is understood to be 
convoluted with the neutrino spectrum $\lambda(E)$. 
Then the slope  becomes less negative as the second term 
dominates over the first, i.e. for $E\sim T$ and 
${\tilde{g}_{\al L}}/{\tilde{g}_{\al R}} <0$, and at the same time 
the width of the neutrino spectrum is narrow enough to resolve 
this term, that is $\lambda(E) \lsim 
|\frac{\tilde{g}_{\al L}}{\tilde{g}_{\al R}}|\frac{m_e}{2 E}$. 
Notice that in the SM case,  ${{g}_{ L}}/{{g}_{R}} >0$ for $\nu_e$ and 
${{g}_{L}}/{{g}_{R}} <0$ for $\nu_\tau$. 
Then the derivative (\ref{deriv}) cannot become positive 
for the {\it continuous} 
$^8$B spectrum. On the other hand, for the mono-energetic $^7$Be neutrinos 
it can  
even for $\eps_{\al R,L} =0$ 
when, for example,  
$P$ is small enough  so that the contribution to the signal 
of $\nu_a$ gets increased and the  role of the interference term 
in the differential cross section gets enhanced. Clearly, by switching 
on the NS couplings this effect can be further emphasized as the 
spectra displayed in Fig.~\ref{p6} show. 
The spectral deformation  can be experimentally probed by sub-dividing the  
energy window considered so far into the two symmetric 
windows which visually emerge from these plots, $\Delta T_1= 3 -4.5$ MeV and   
$\Delta T_2= 4.5 - 6$ MeV, and comparing the respective number of events 
$N_1$ and $N_2$.\footnote{
Our choice of the energy windows $\Delta T_1, \Delta T_2$ is 
somehow arbitrary. The choice should be better motivated  
by background considerations.}
For this purpose, we have introduced the asymmetry parameter $\delta$:
\be{delta}
\delta =\dfrac{ N_1 -N_2}{N_1 + N_2}  .
\ee
From the upper plots of Fig.~\ref{p6}, we can see 
that the asymmetry $\delta$ is 0.046 for $\eps_{R}=0$ and 
 further diminishes 
for negative values of $\eps_{e R }$, while  
it increases for positive $\eps_{e R }$, up to 11\% for $\eps_{e R }=0.4$. 
The same behaviour occurs when $\eps_{\tau R }$ are non-vanishing, 
though both the increasing and the decreasing is less dramatic. 
Finally, the lower panel of Fig.~\ref{p6}  shows the correlation between 
the total signal $N_{\rm tot}=N_1+N_2$ and the event difference $N_1-N_2$ 
(i.e. the asymmetry $\delta$) for different values of $\eps_{e R}$ 
(solid lines) and   $\eps_{\tau R}$ (dashed lines) and for comparison 
with the SM case (dotted line). 
These curves demonstrate that the joint measurement of $N_{\rm tot}$ and 
$N_1-N_2$ could reveal the  presence of neutrino NS 
interactions.  
In the SM case, the asymmetry is almost independent of $N_{\rm tot}$  
in the interval 
$N_{\rm tot}= 17- 38$, (corresponding to the survival probability 
in the range $P=0.3-1$) as it varies from $\delta \simeq 0.044$ to 
$\delta \simeq 0.046$, and it approaches the minimal 
value $\delta \simeq 0.037$  for $N_{\rm tot}= 8.3$  ($P=0$), when the 
signal is only contributed  by the $\nu_{\mu,\tau} e$ scattering.
For non-vanishing $\eps_{e R}$, 
the variation of the parameter $\delta$     
with respect to $N_{\rm tot}$ is more pronounced. Namely, in the interval 
$N_{\rm tot}= 17- 38$ (corresponding incidentally to the same interval 
$P=0.3-1$)   
the asymmetry varies from $0.10$ to 0.13 and for smaller 
$N_{\rm tot}$ it sharply decreases.    
This asymmetry parameter or, equivalently, the measurement 
of the difference $N_1-N_2$ should be statistically significant 
for one-year of data 
taking. For example, by taking $P=0.5$, we expect in the SM case 
$N_{\rm tot}\simeq 8400$ per year 
and a difference of events $N_1- N_2 \simeq 380$ which is larger than 
the statistical fluctuation.
Analogously for $\eps_{e R}=0.4$, we find the same $N_{\rm tot}$ 
per year, but an even more statistically significant
  difference of events, $N_1- N_2 \simeq 940$. 
The cases with non-vanishing $\eps_{\tau R}$ show 
a less dramatic effect with 
respect to the SM case, unless $\eps_{\tau R}\gsim 0.2$.    
Notice, that it may be essential to make use of  
the interplay between the measurement of the 
the energy spectrum $S(T)$ and the corresponding asymmetry 
$\delta$  to better disentangle 
scenarios with different NS couplings 
(either $\eps_{e R}$ or $\eps_{\tau R}$ or both  
non zero).

\section{Conclusions}

Extensions of the Standard Model generally include new neutral current 
interactions that can be flavour changing as well as flavour conserving. 
In this respect, the recent 
 NuTeV data are very encouraging and should be further considered.

In this paper, we have re-examined the implications for solar neutrino 
detection of non-standard flavour conserving 
interactions of neutrinos with the electron. 
In particular, we have payed special attention to  the case in which 
$\nu_\tau$ and $\nu_e$ have non-standard couplings with the  electron.
In the light of the atmospheric neutrino data pointing to a large mixing 
between $\nu_\mu$ and $\nu_\tau$, the solar neutrino deficit can now 
be regarded as the `conversion' of $\sim$ 65\% of solar $\nu_e$'s into 
an equal amount of $\nu_\mu$ and $\nu_\tau$. 
We have found that 
 present solar neutrino experiments  show a complementary  sensitivity 
in the parameter space 
with respect to that 
achieved  by laboratory neutrino experiments. 
We have so tried to reverse the point of view 
by looking for some signature of such NS interactions within 
the allowed parameter space.
We have  demonstrated that 
the allowed range for these neutrino non-standard interactions 
could be tested by Borexino experiment, aimed to detect the 
monochromatic $^7$Be-line.  
Indeed, neutrino NS interactions could manifest especially   
through  an unexpected  spectral deformation and in some  cases 
with  an `anomalous' increase of the number of total events. 
Therefore, the Borexino experiment can be a unique tool to test 
solar neutrinos as well as the Standard Model itself.
Now we would like to comment on an issue related to 
future low-energy neutrino experiments. 
For the sake of simplicity, we consider only $^7$Be neutrinos.   
Then we consider the following function, normalized 
to the SSM expectation:
\be{tau-shape}
F(T)
= \dfrac{ c^2_{23} \dfrac{d\tilde{\sigma}_{\nu_\mu}(E,T)}{dT} + 
 s^2_{23} 
\dfrac{d\tilde{\sigma}_{\nu_\tau}(E,T)}{dT}}
{\dfrac{d\tilde{\sigma}_{\nu_e}(E,T)}{dT}}  .
\ee
This  observable $F(T)$  can be 
in fact built up by subtracting   from the 
energy distribution $S(T)$ measured at Borexino the survived $\nu_e$ 
contribution as  `directly' measured 
  by a low-energy neutrino 
experiment such as LENS -- sensitive {\it only} to the $\nu_e$ flavour 
\cite{LENS}. In this way we would more directly test 
 the $\nu_\mu, \nu_\tau$ contribution to the 
energy distribution and so test $\nu_\tau$ NS interactions. 
This complementary role between these two kinds of experiments 
is analogous to that presently 
played by SK and SNO for the high-energy $^8$B neutrinos.

Finally, we would like to comment about the possibility, which is 
usually neglected,  to solve the solar neutrino deficit with matter 
oscillations induced  by 
flavour-changing and flavour-conserving 
NS interactions of neutrinos with electrons (with massless neutrinos) 
\cite{BPW}.  Indeed, 
also  such  flavour-changing NS 
interactions (the relevant flavour-changing parameters would be 
$\eps_{e \beta V},~ \beta=\mu,\tau$)  are not strongly constrained
\cite{az3} and, jointly with the flavour-diagonal NS ones, can 
give rise to a sizeable mixing angle in matter.
Therefore, it could be worth re-considering 
such conversion mechanism.\footnote{
This possibility was discussed in \cite{GNHP}; however, the 
effect of neutrino NS interactions with electrons was disregarded 
in the detection cross section.}

\vspace{0.3cm}
{\bf Acknowledgments} 
We  thank  
L.~Girlanda and P.~Ronchese for `graphics' assistance. 
The work of Z.~B. was in part supported by the MURST Research Grant 
"Astroparticle Physics" and that of A.~R.  
by the European Union under the contracts 
HPRN-CT-2000-00148 (Across the Energy Frontier).

\newpage
\begin{figure}[p]
\vskip -3.cm
\hglue -0.5cm
\epsfig{file=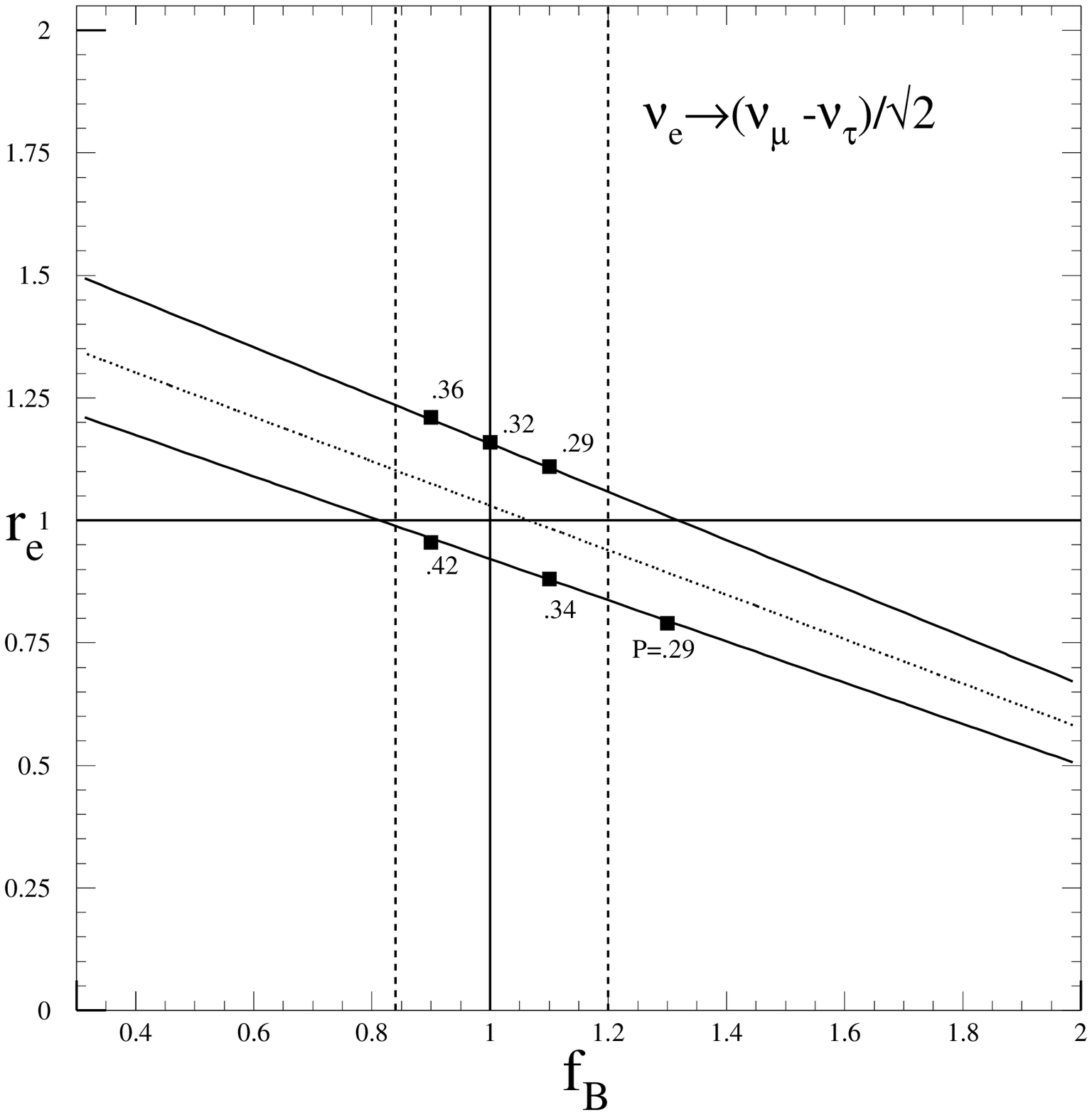,height=7.8cm,width= 8.6cm}
\vglue - 7.8cm
\hglue  7.5cm
\epsfig{file=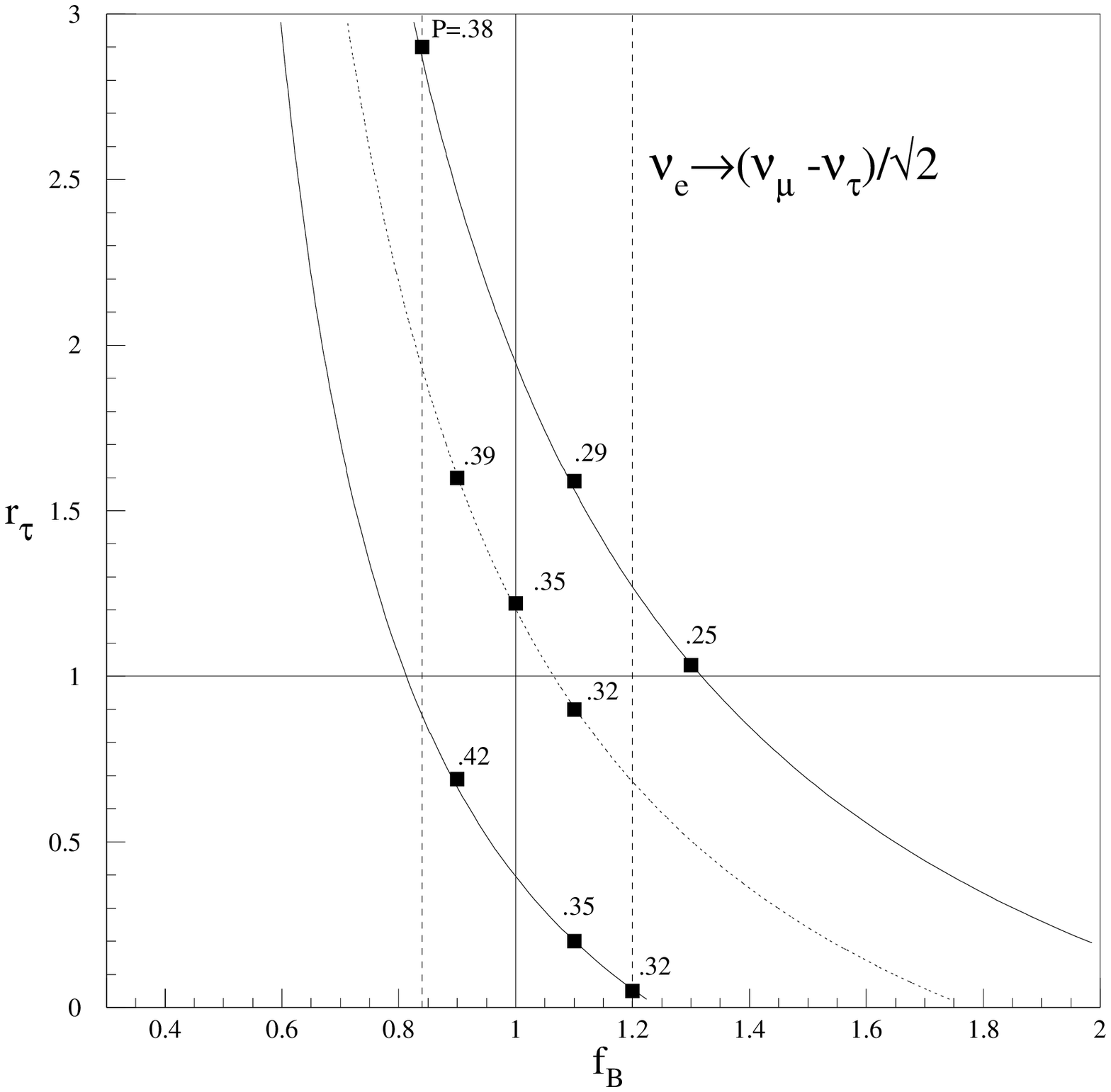,height=7.8cm,width= 8.6cm}
\vskip -0.5cm
\caption{\small
Analysis of the SK and SNO data for the conversion
$\nu_e\to \nu_a = c_{23}\nu_\mu -s_{23}\nu_\tau$.
In the left panel only $\nu_e$ is considered to have the
NS couplings with electrons ($r=r_e$),
while in the right panel only $\nu_\tau$ ($r=r_\tau$),
and  maximal mixing is assumed, $\theta_{23}=45^\circ$.
The contours of the 1$\sigma$ allowed region are shown by
solid curves (the dotted lines in
the middle refers to the central values of SK and SNO data).
The range of $f_B$ allowed by the SSM uncertainties
is delimited by vertical dashed lines
($f_B=1$ corresponds to the reference SSM model).
As a guideline for the reader we have drawn (black squares)
some points for the $\nu_e$ survival probability which for given
$f_B$ is fixed by the SNO signal, $P=Z_{\rm SNO}/f_B$.
}
\label{p2}
\end{figure}

\vskip -7.cm
\begin{figure}[hb]
\vskip -3.cm
\hglue -0.6cm
\epsfig{file=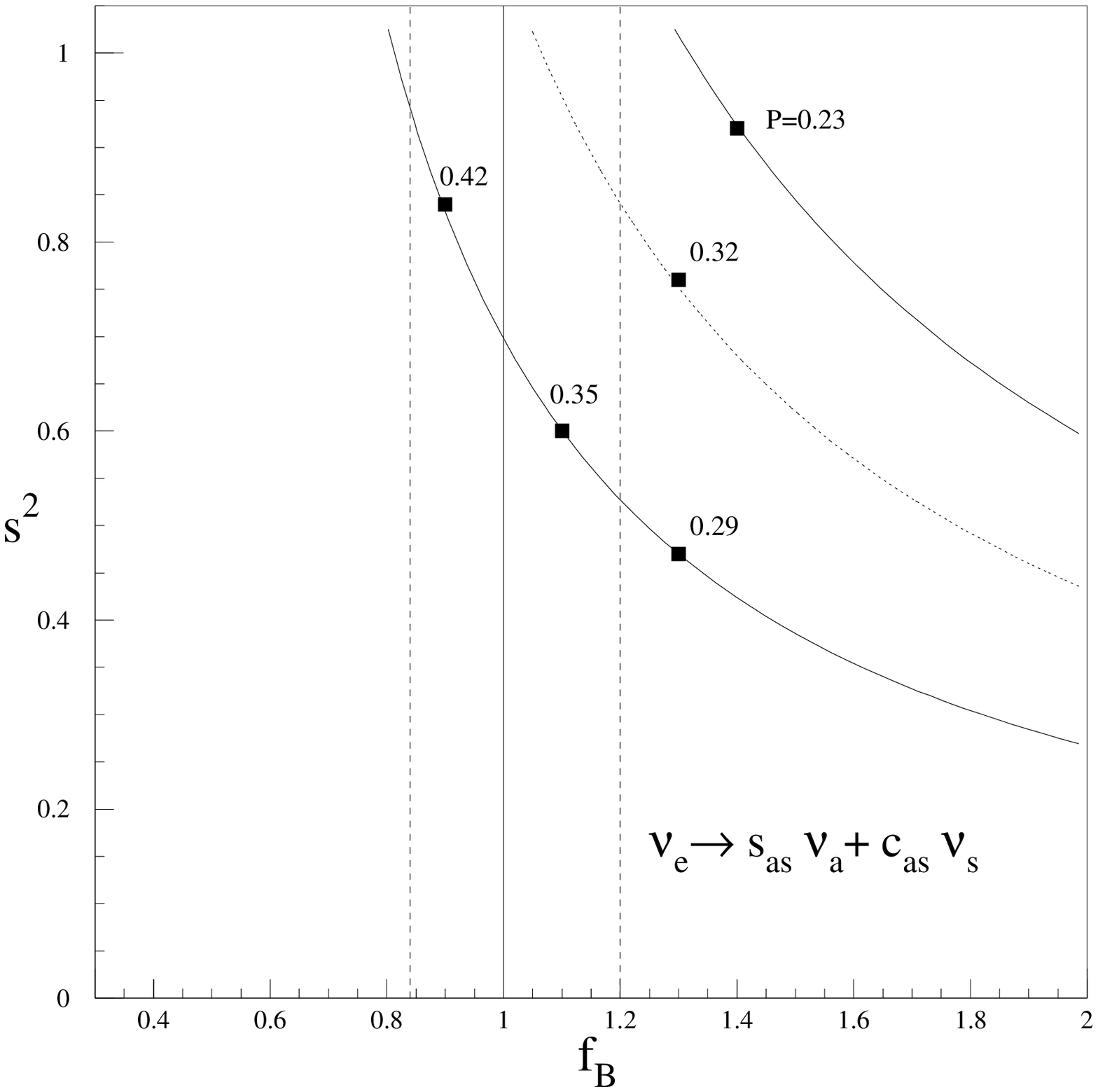,height=7.8cm,width= 8.6cm}
\vglue - 7.8cm
\hglue  7.5cm
\epsfig{file=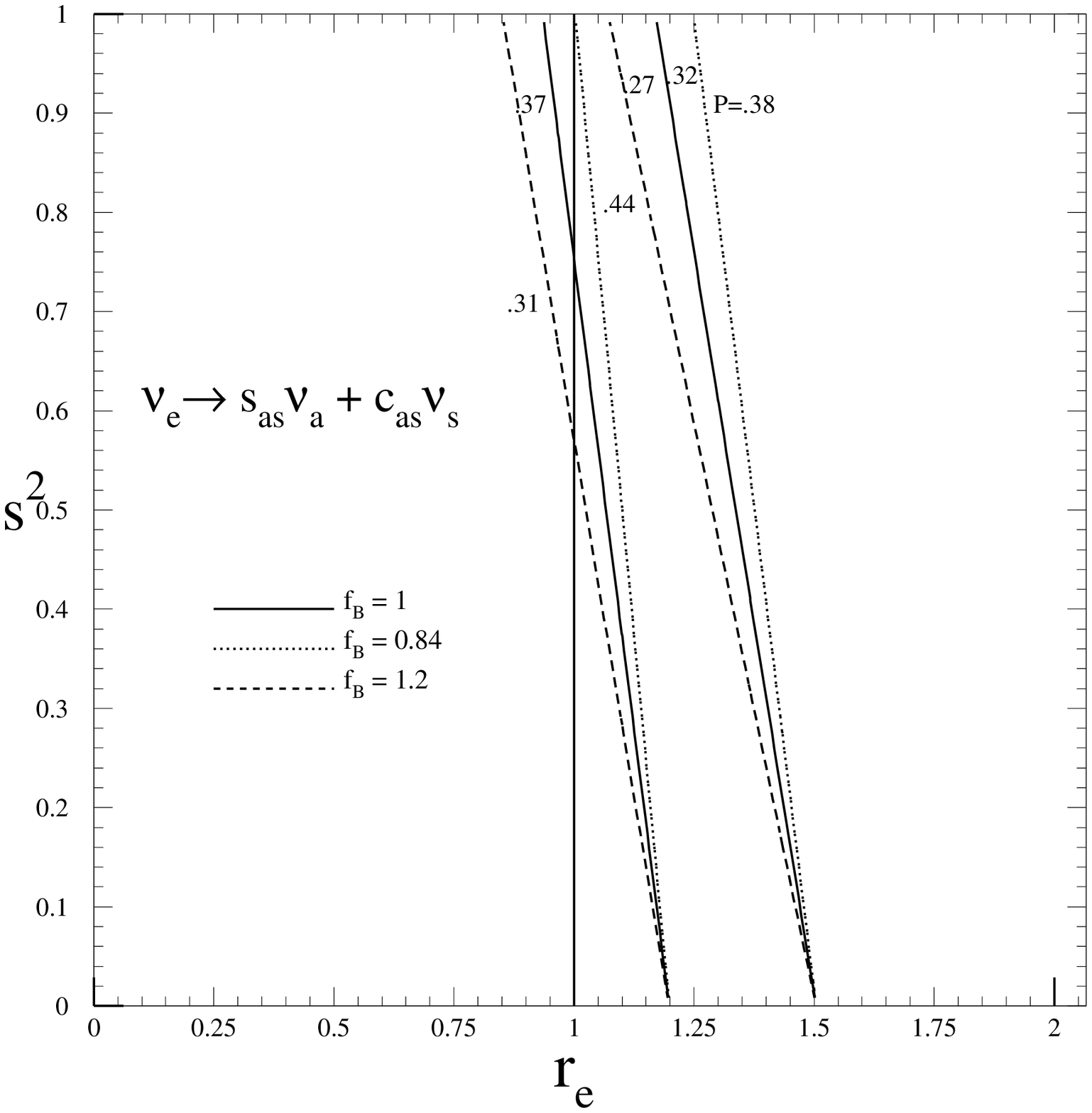,height=7.8cm,width= 8.6cm}
\vskip -0.6cm
\caption{\small
Analysis of SK and SNO data regarding the conversion
$\nu_e\to \nu^\prime = s_{as} \nu_a + c_{as}\nu_s$
($\nu_s$ is a sterile neutrino).
On the left panel the standard case ($\eps_{R,L} =0$) is illustrated
in the plane scanned by the
paramemers $f_B$ and $s^2=s^2_{as}$;
the solid contours delimit the region allowed by
the SK/SNO signals (\ref{Zss1}) at 1$\sigma$,
the dotted curve in the middle corresponds
to the central values of $Z_{\rm SK}$ and $Z_{\rm SNO}$.
The values of $P$ are indicated by black squares.
On the right panel the effect of NS interactions of $\nu_e$ is accounted
in the plane ($r_e,s^2$).
The `oblique' solid lines delimit  the parameter space
allowed by SK/SNO for $f_B=1$, while the effect of
varying $f_B$ within $0.84-1.2$ is accounted
by dot- and dash-lines respectively. The corresponding
survival probabilities $P=Z_{\rm SNO}/f_B$ are also
indicated along these lines.
}
\label{p1}
\end{figure}

\vskip -1cm
\begin{figure}[p]
\vskip -2.cm
\hglue -0.5cm
\epsfig{file=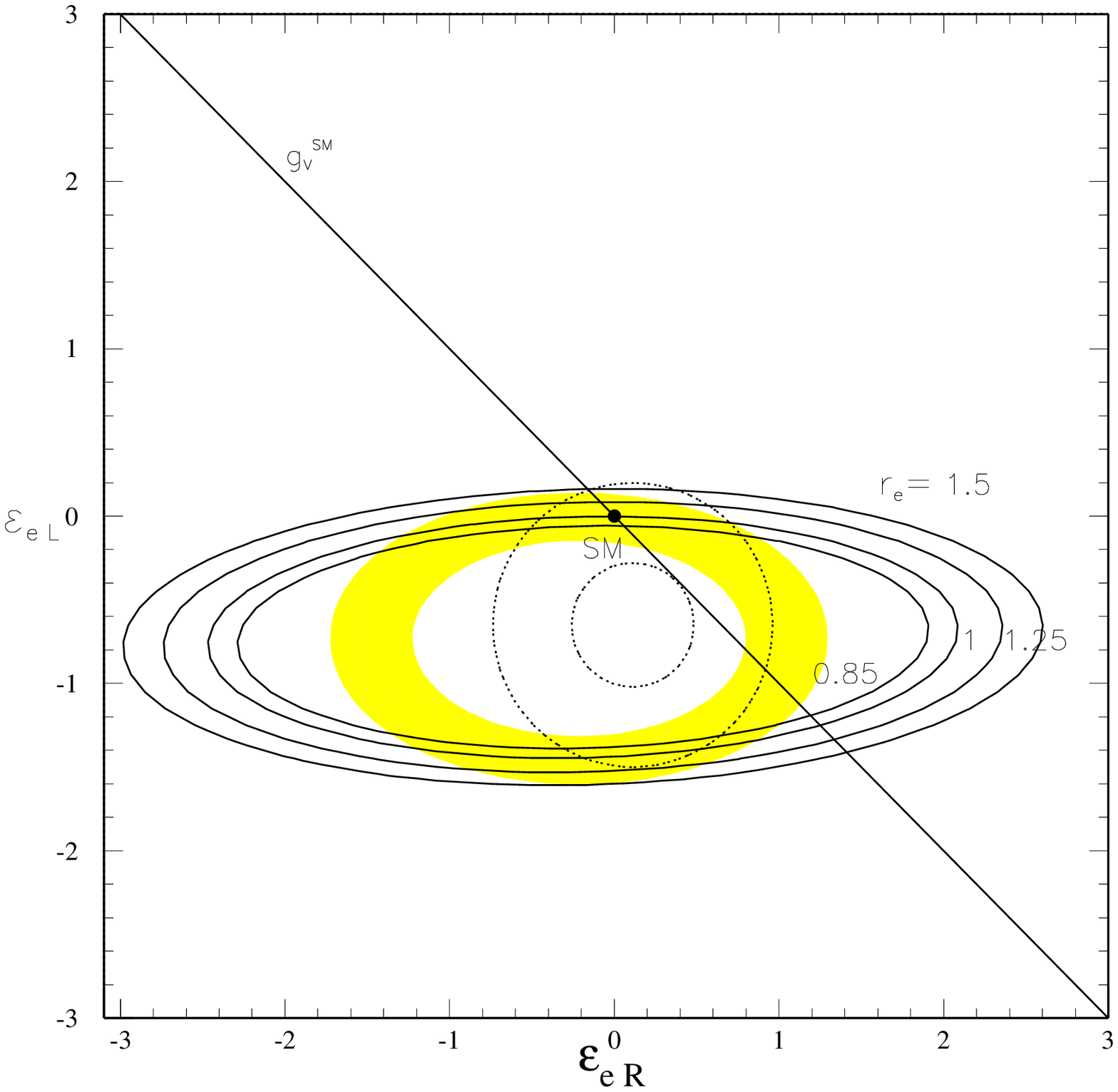,height=8.6cm,width= 8.6cm}
\vglue -8.6cm
\hglue 7.5cm
\epsfig{file=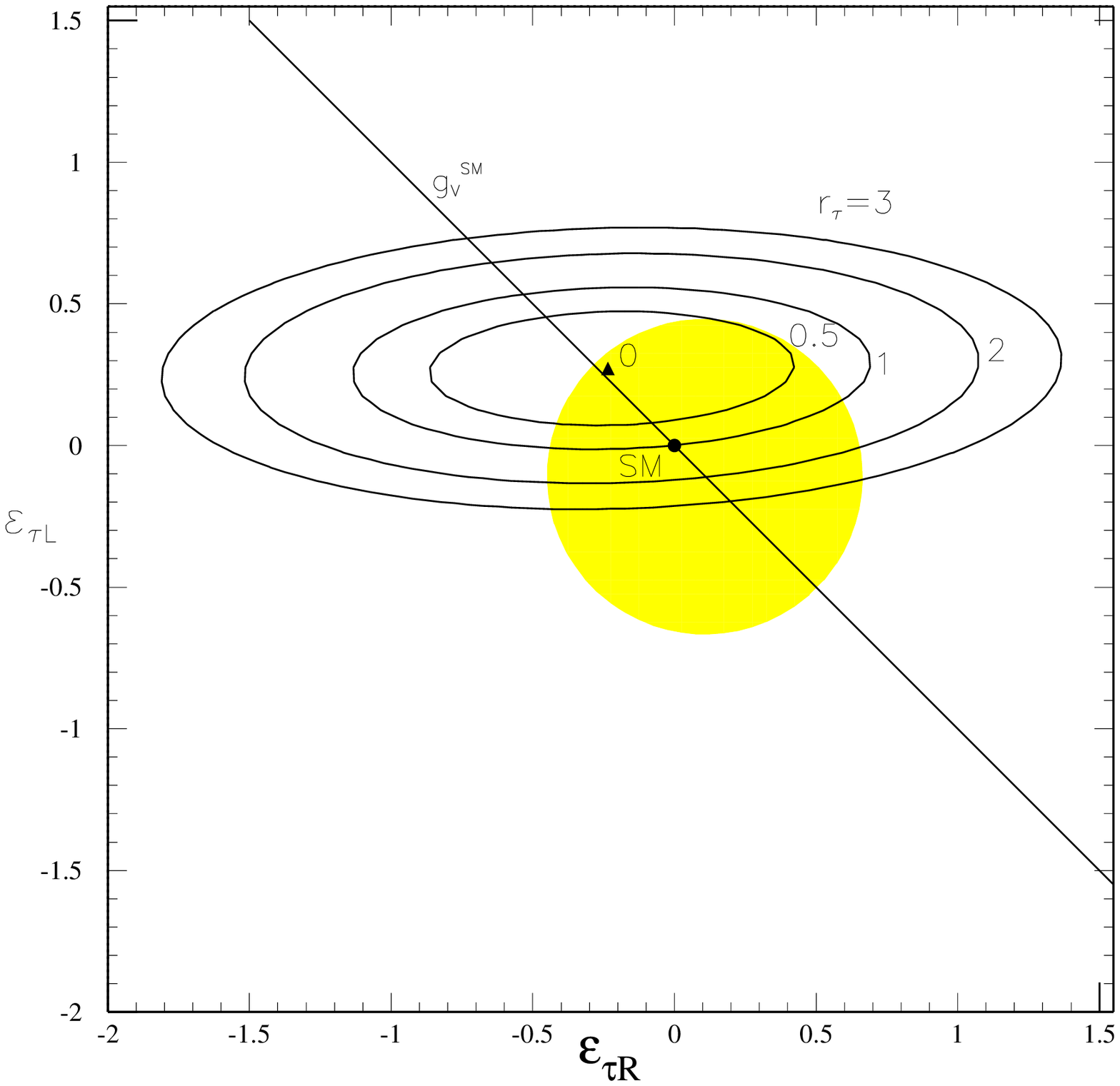,height=8.6cm,width= 8.6cm}
\vskip -0.3cm
\caption{\small
Left panel: 
experimental bounds on the NS interactions of $\nu_e$.
The shaded area and the annulus (delimited by dotted lines)
represent the parameter spaces for $\eps_{eR}, \eps_{eL}$
allowed 
by data on $\nu_e e$ 
elastic scattering cross section and
$e^+e^-\to \nu\bar{\nu}\gamma$ cross-section, respectively, 
at 99$\%$ C.L. \cite{az3}.
Right panel:  bounds on the  NS interactions of
$\nu_\tau$. The shaded area encloses  the parameter space for
$\eps_{\tau R}, \eps_{\tau L}$ allowed by data  the
$e^+e^-\to \nu\bar{\nu}\gamma$ cross-section at $99\%$ C.L. 
\cite{az3}. On both panels, 
the contours for different values of
$r_e$ and $r_\tau$ in the SK signal are also shown 
(solid curves).
The SM case itself ($\eps_{R,L}=0$)
is indicated by filled circles, and $g^{SM}_V$ labels the lines
along which $\eps_{e(\tau) V} =0$.
}
\label{p3}
\end{figure}

\begin{figure}[h]
\hglue -0.7cm
\epsfig{file=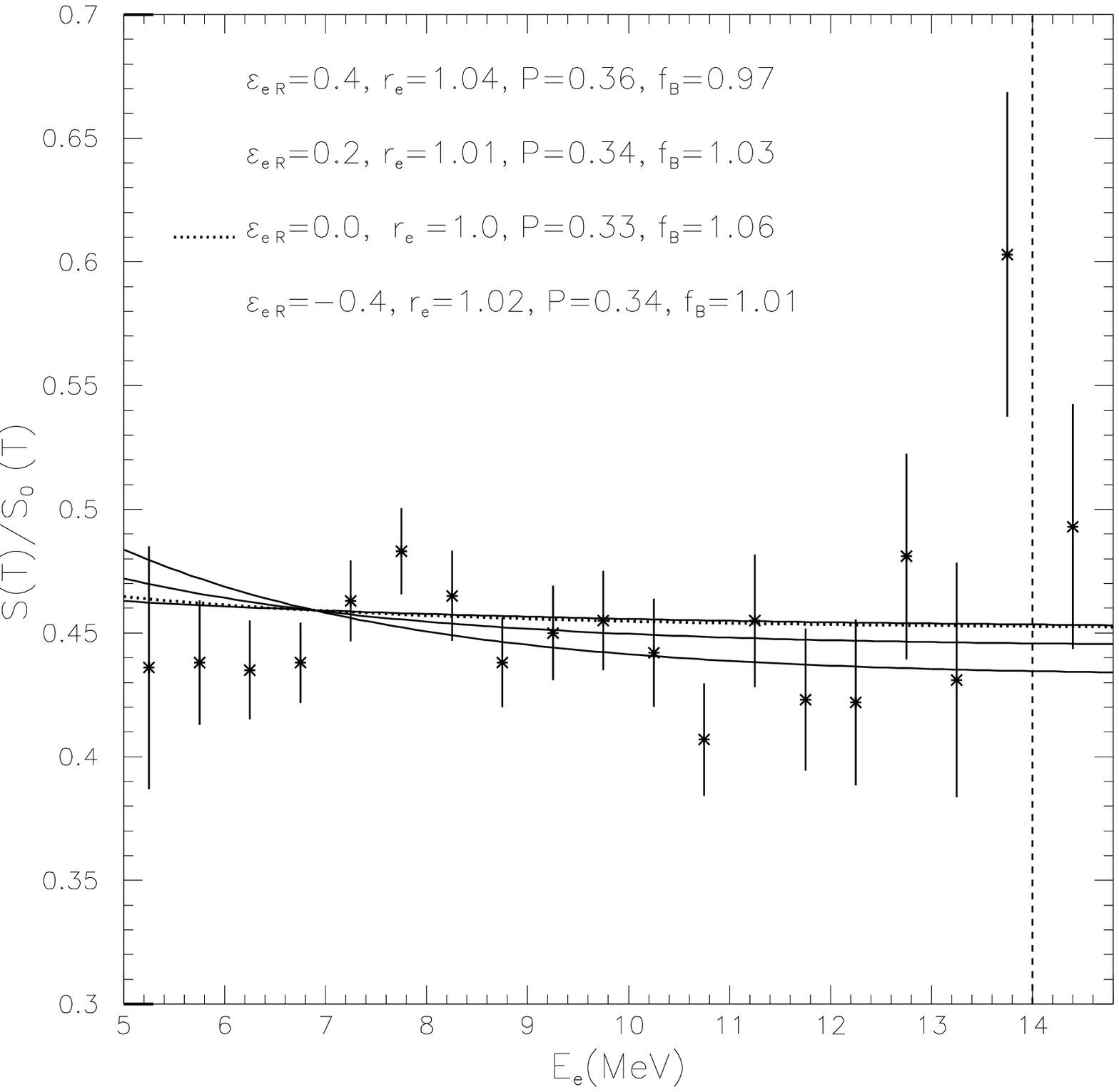,height=7.8cm,width= 8.6cm} 
\vglue -7.8cm
\hglue 7.5cm  
\epsfig{file=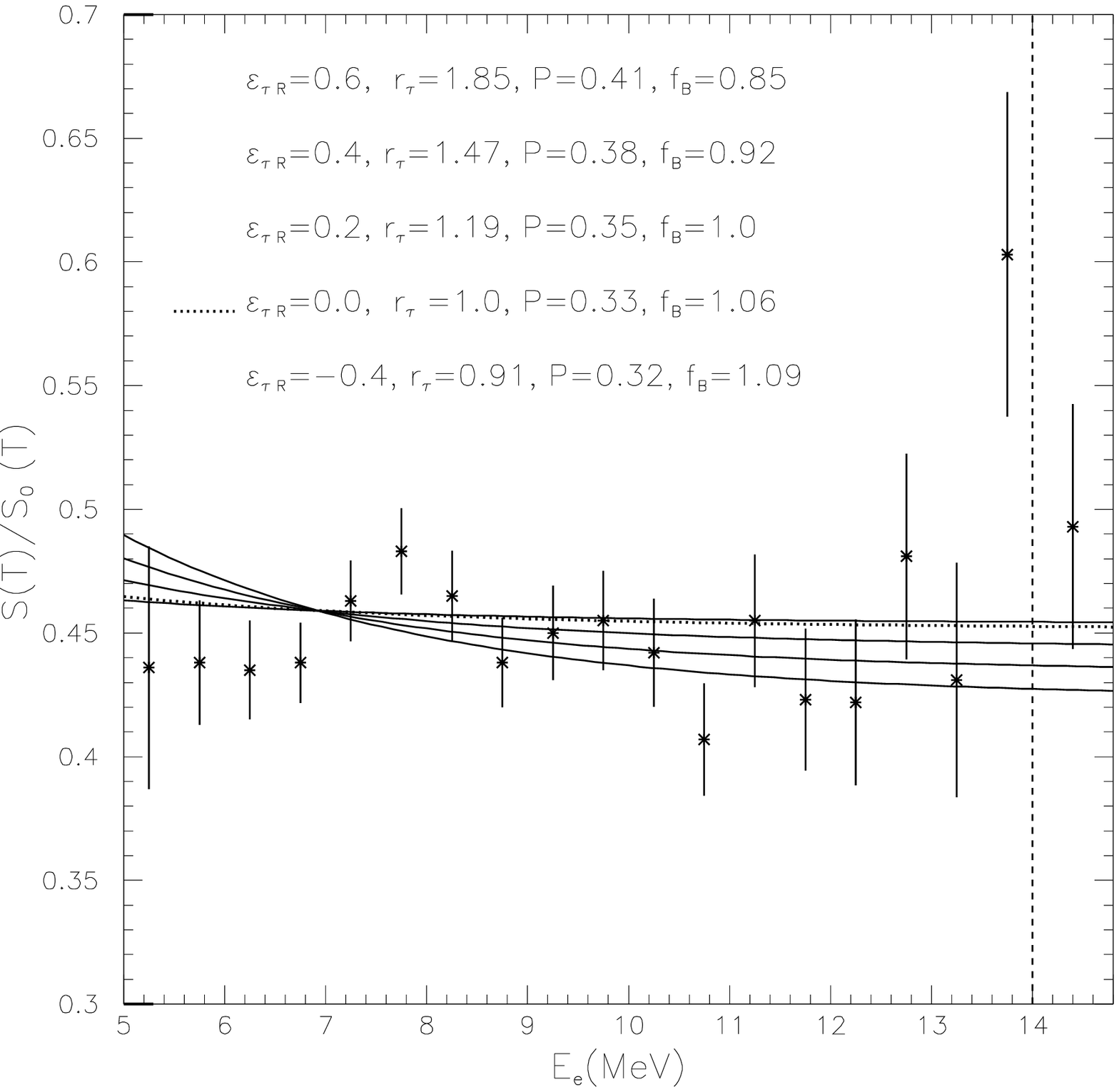,height=7.8cm,width= 8.6cm}
\caption{\small
Left panel:
SK energy spectra
(divided by  the SSM expectation)
versus the recoil electron energy $E_e$ for different
values of $\eps_{e R}$ in the allowed range
($\eps_{e R}=0.4,0.2,-0.2,-0.4$), and $\eps_{eL}=0$.
The SK experimental data (with the corresponding error) are also
shown \cite{SK}.
Right panel: the same but for different values of
$\eps_{\tau R}$ ($\eps_{\tau L}=0$).
All spectra  are normalized to the central value of
$Z_{SK}$ through the constraints (see eq.~(\ref{fB-P}))  imposed on
$P$ and $f_B$ by the experimental central values of $Z_{SK}$ and $Z_{SNO}$
for given $\eps_{e R} (\eps_{\tau R})$  i.e. $r_e (r_\tau)$ (as indicated).
The inferred  values of $P, f_B$ are also  shown. 
}
\label{p4}
\end{figure}

\begin{figure}[p]
\hglue -0.8cm
\epsfig{file=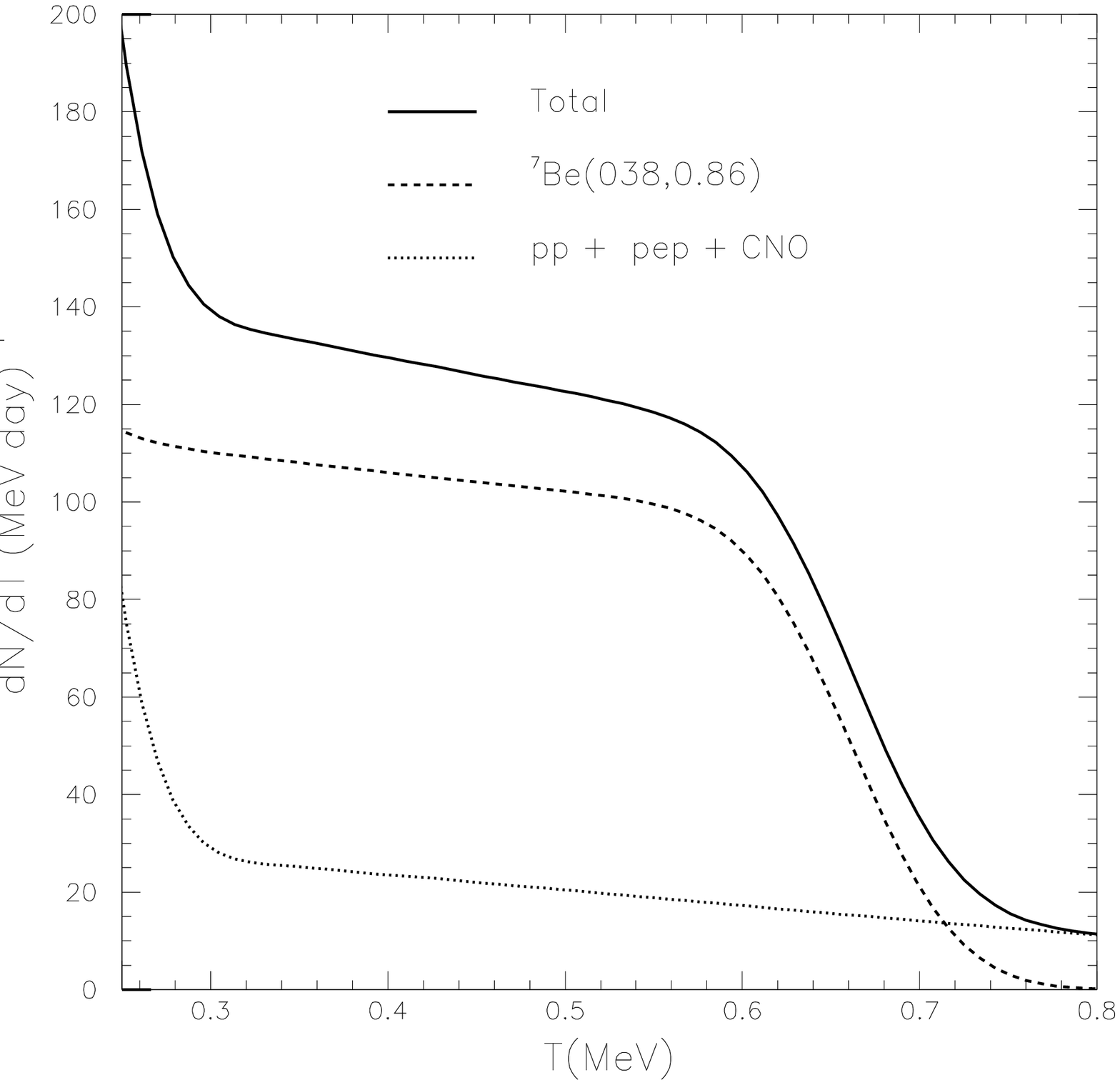,height=9.cm,width= 8.6cm}
\vglue -9.cm
\hglue 7.6cm
\epsfig{file=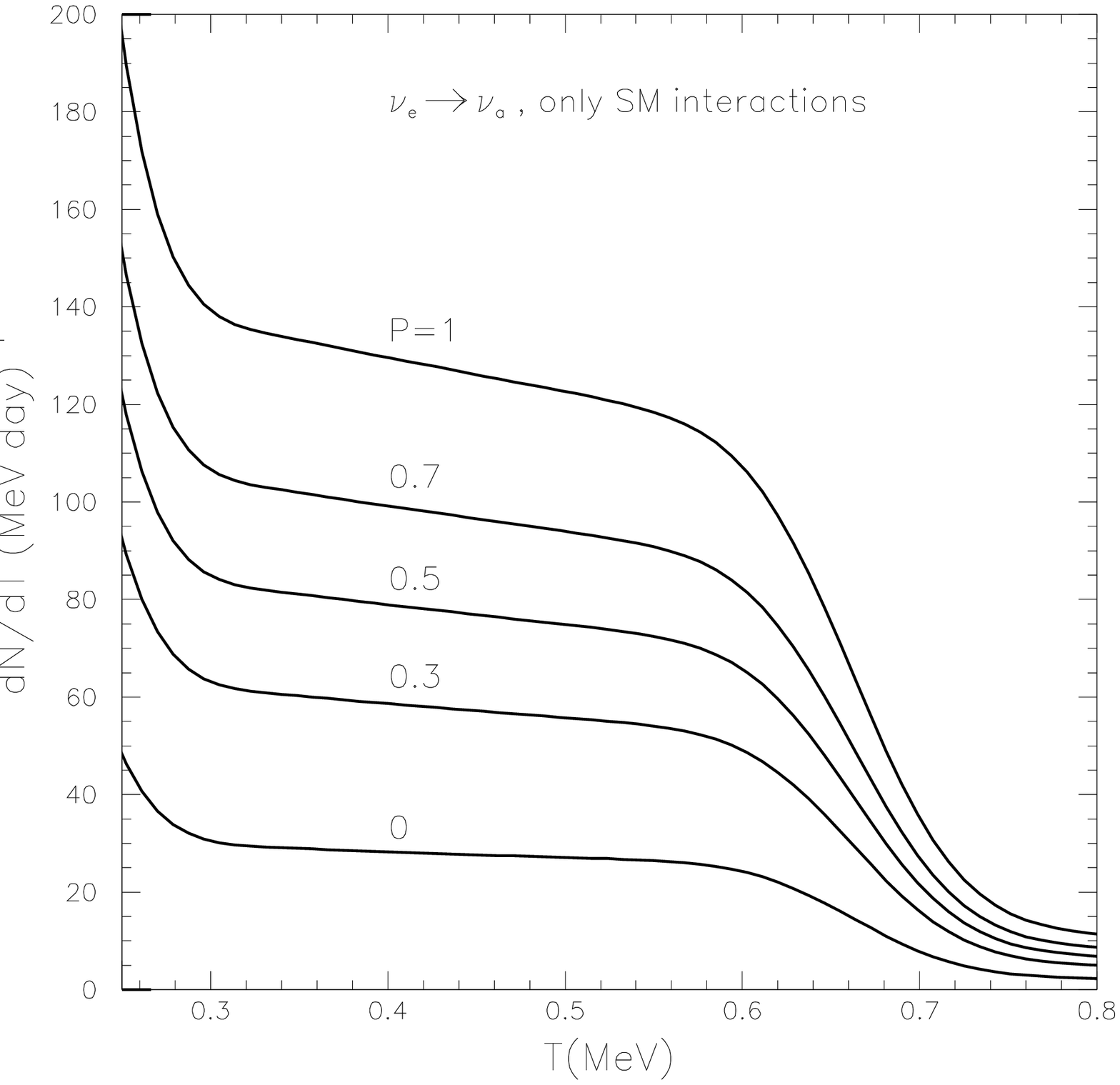,height=9.cm,width= 8.6cm}
\vglue 0.9cm
\hglue -0.8cm
\epsfig{file=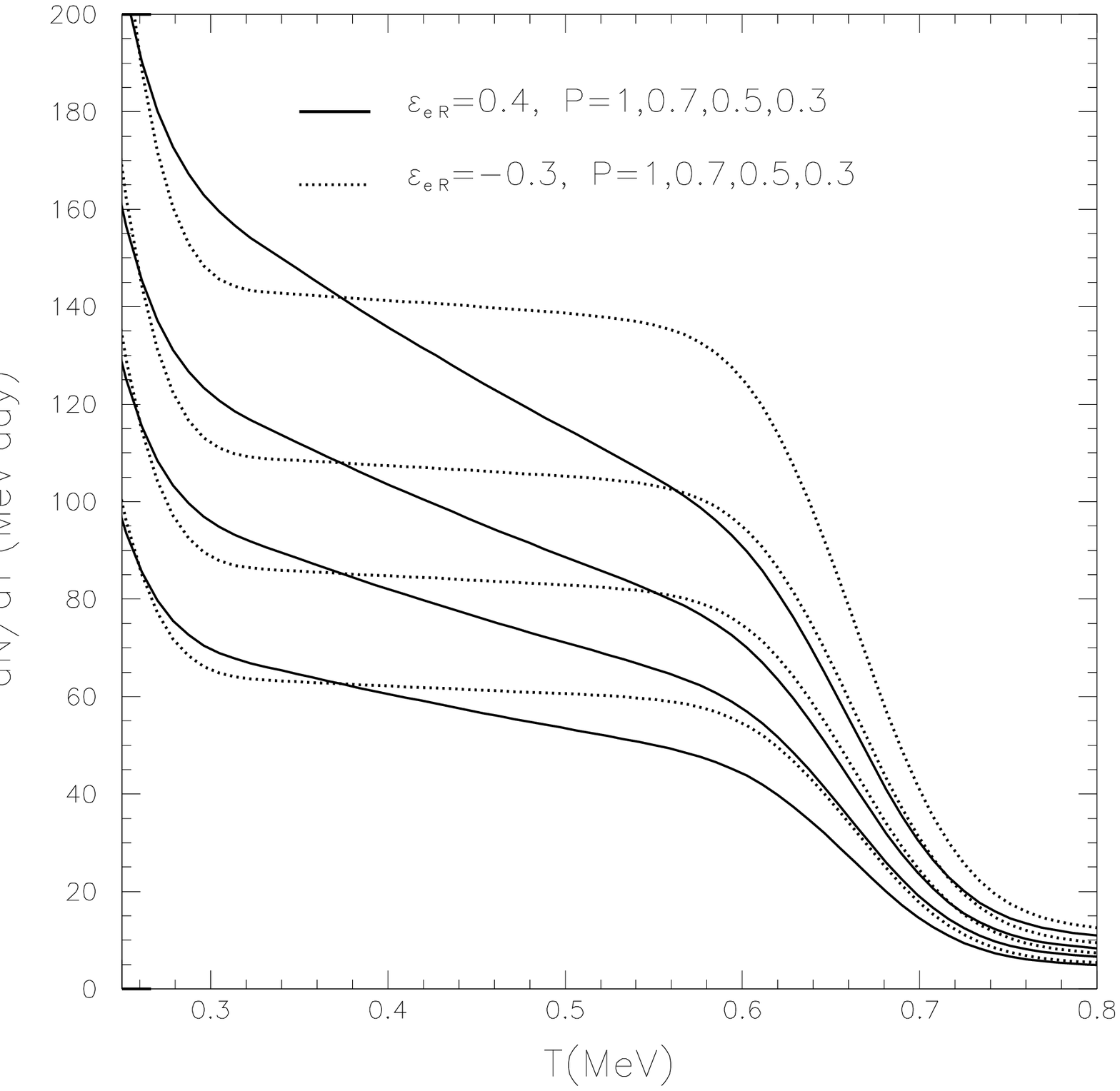,height=8.5cm,width= 8.6cm}
\vglue -8.5cm
\hglue 7.6cm \epsfig{file=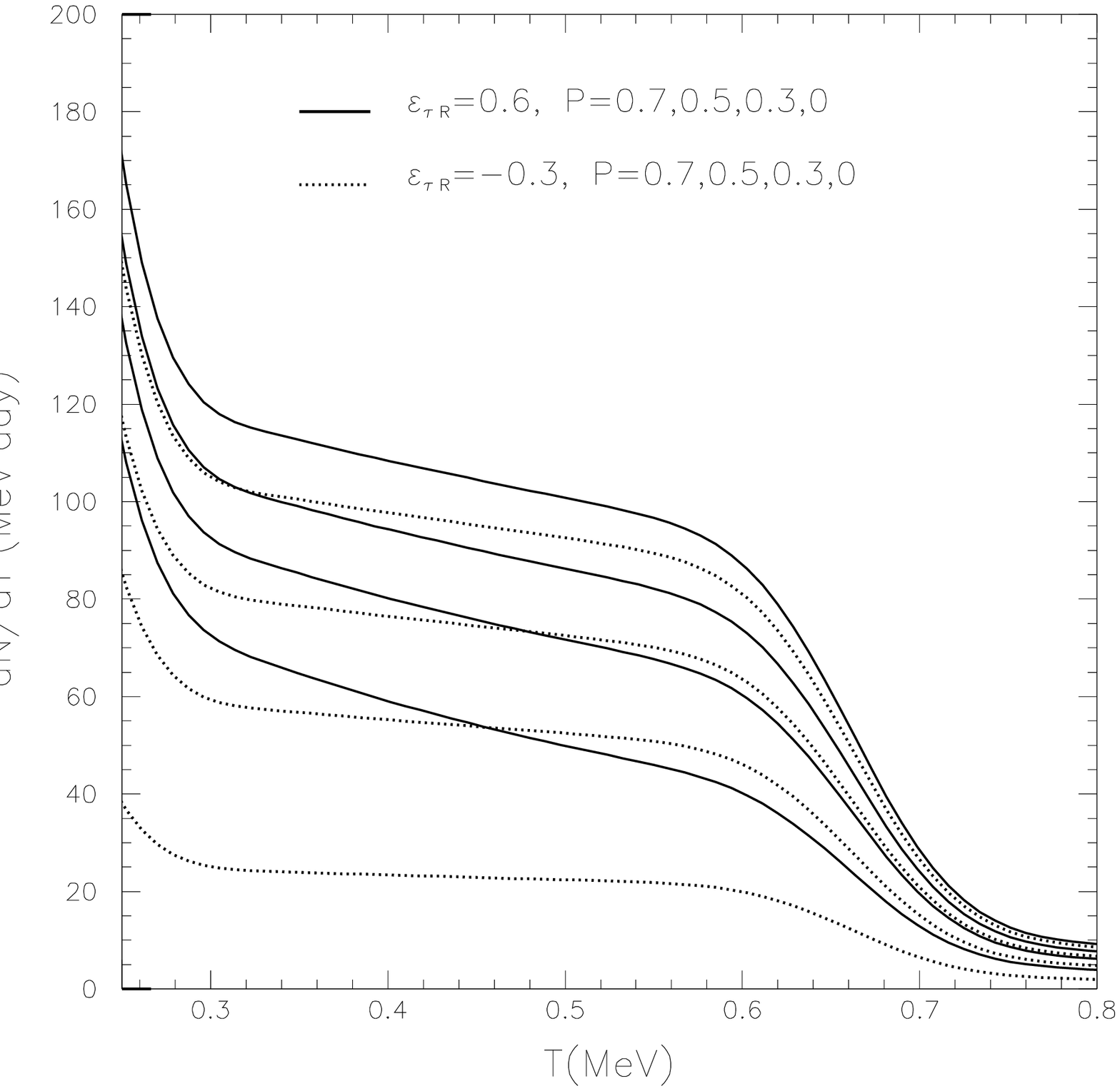,height=8.5cm,width=8.6cm}
\caption{\small   
Upper left panel: 
the event distribution in Borexino experiment
versus the recoil electron kinetic energy $T$, as expected
for the SSM predicted fluxes ($P=1$)
and with no extra interactions ($\eps_{L,R}=0$).
The dotted curve shows the signal due to only
$^7$Be neutrinos ($E=0.38, 0.86$ MeV),
while the dashed curve indicates the contribution
from all other relevant solar $\nu_e$ sources
($pp$, $pep$, CNO etc.).
The total expected signal is shown by the solid curve.
Upper right panel: 
the effect on the energy distribution of the events due to
the neutrino conversion $\nu_e\to \nu_a$ for
different survival probabilities $P$.
Lower panels show how the event distribution gets modified  
for non-zero $\eps_{e R}$ (left) or $\eps_{\tau R}$ (right) 
for different suervival probabilities 
($\eps_{e,\tau L}=0$ is understood everywhere).
}
\label{p5}
\vskip -0.1cm
\end{figure}

\begin{figure}[p]
\hglue -0.5cm
\epsfig{file=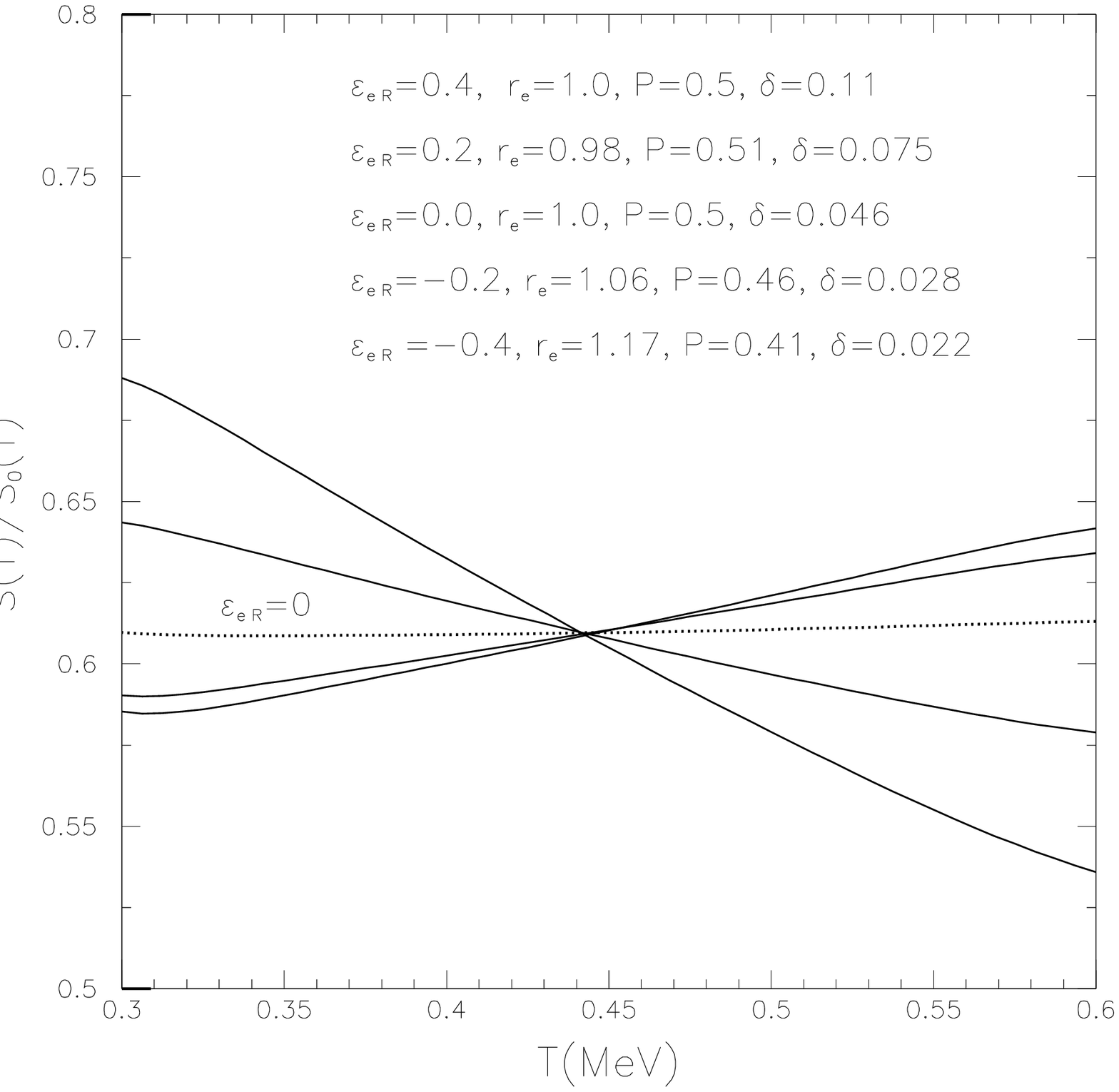,height=7.8cm,width= 8.6cm}
\vglue - 7.8cm
\hglue  7.5cm
\epsfig{file=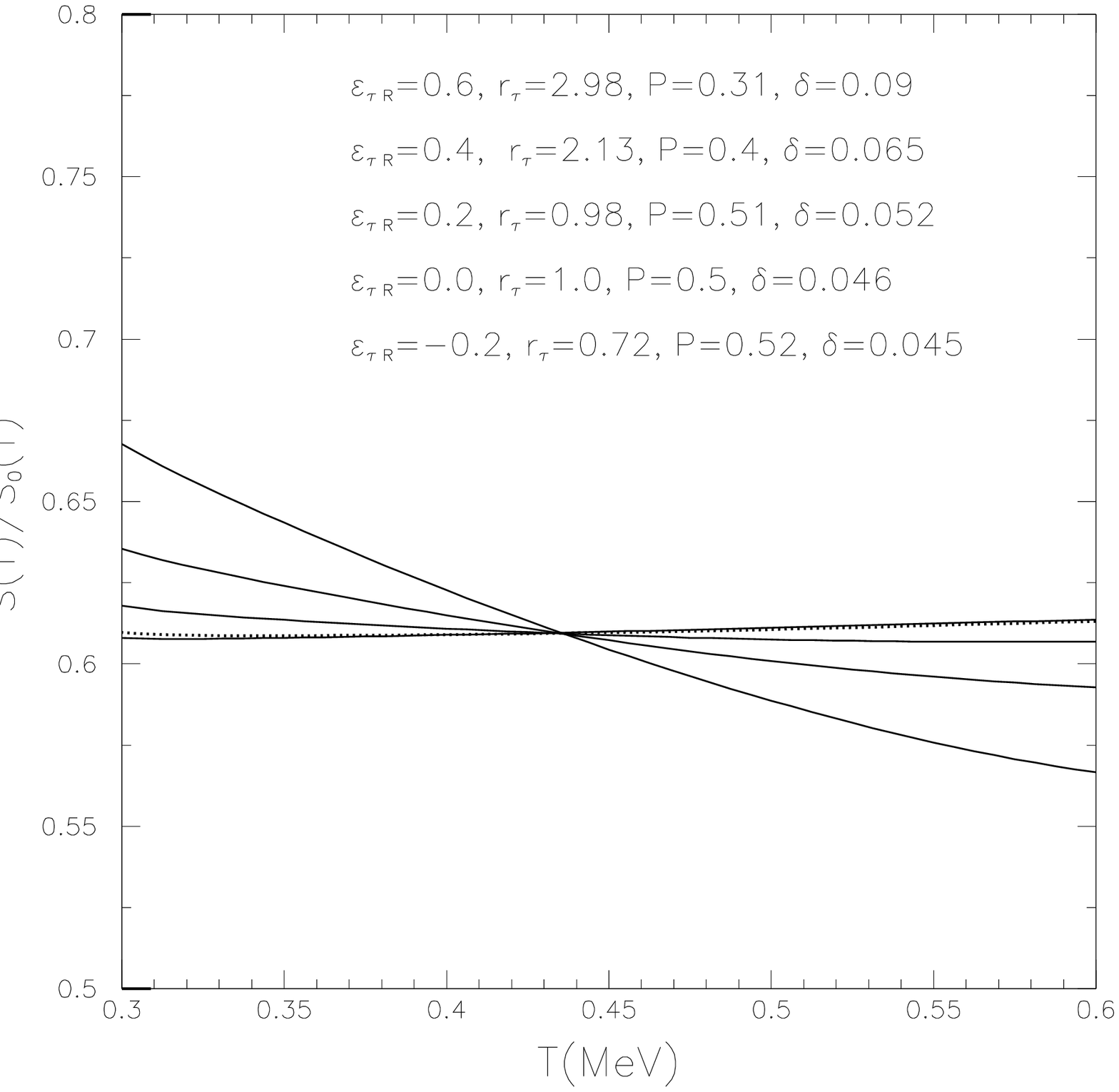,height=7.8cm,width= 8.6cm}
\hglue  3.8cm
\epsfig{file=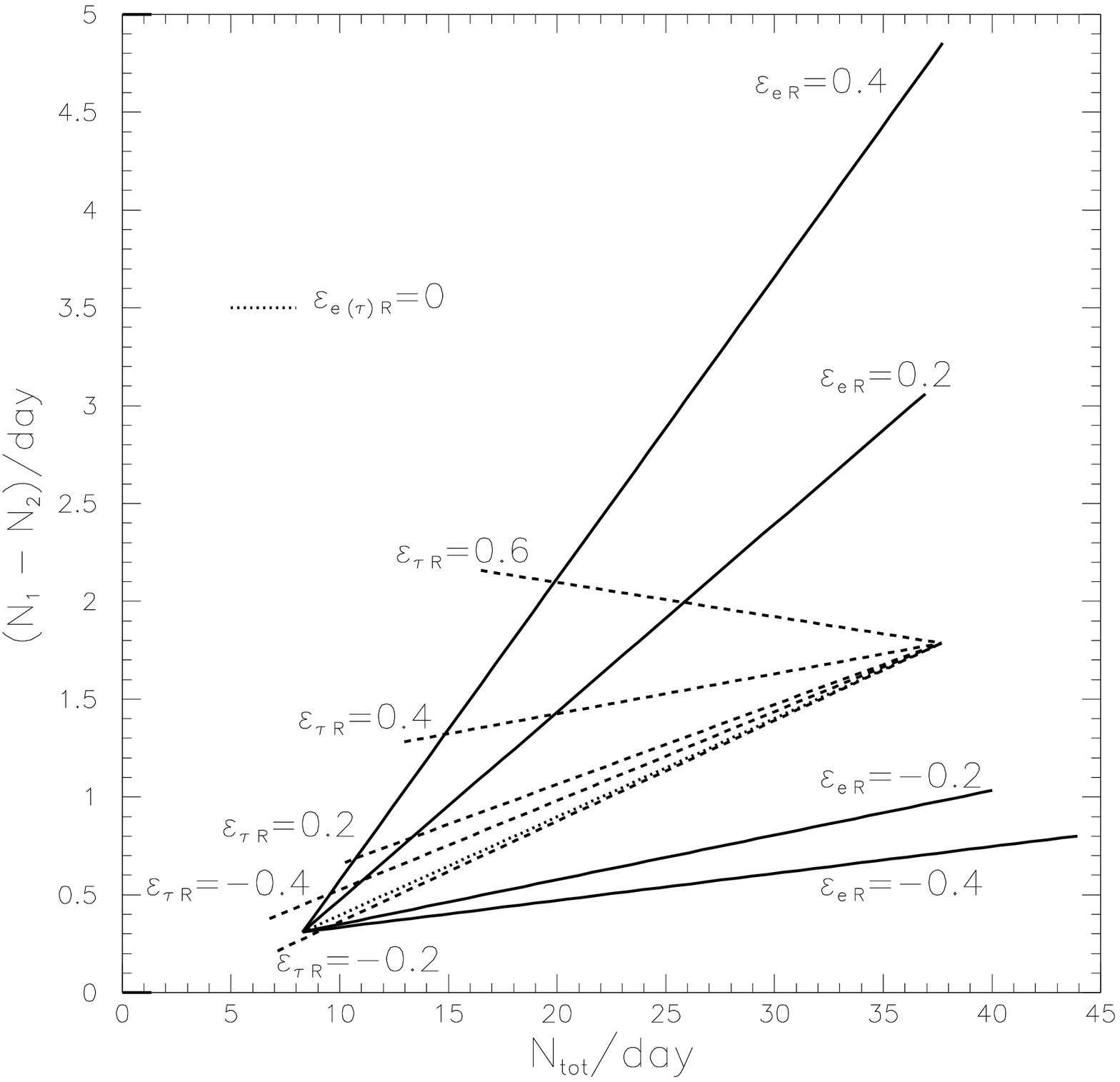,height=7.8cm,width= 8.6cm}
\vskip -0.5cm
\caption{\small 
Upper  panels: $S(T)/S_0(T)$ is the ratio of the  electron energy spectrum 
for non-vanishing  
$\eps_{e R}$ (left panel) and $\eps_{\tau R}$ (right panel) 
with that 
expected from the SSM (and $\eps_{e, \tau R}=0)$ at Borexino.
All curves are normalized in  such a way that the total number of
events  be $N_{\rm tot}=23$ per day in the interval
$T=0.3 - 0.6$ MeV (obtained with $P=0.5$ in the SM case, $\eps_{R}=0$).
The corresponding  values of $r_e$ (or $r_\tau$), 
$P$ and the asymmetry $\delta$
are also indicated.
Lower panel: 
The  correlation between the expected $N_{\rm tot}$ 
and the difference of events $N_1-N_2$  
in the energy windows 
$\Delta T_1= (0.3-0.45)$ MeV, $\Delta T_2= (0.45-0.6)$ MeV, 
for several 
values of NS couplings $\eps_{e R}$ (solid lines) and $\eps_{\tau R}$  
(dashed lines). The SM case ($\eps_{e, \tau R}= 0$) is depicted by 
dotted line. 
For $\eps_{e R}\neq 0$, the variation of $N_{\rm tot}$ 
from $\simeq 8.3$ to $\simeq 38 - 44$ 
is due to the corresponding change of the   
survival probability $P$ from 0 to 1. 
For $\eps_{\tau  R}\neq 0$ and  $P=0$, 
$N_{\rm tot}$ varies according to $\eps_{\tau R}$ -- 
e.g.  $N_{\rm tot} \simeq 13$ or 10 for $\eps_{\tau R} =0.6$ or 0.2, 
respectively. In this case, all curves 
clearly converge to $N_{\rm tot} \simeq 38$ 
for $P=1$. 
}
\label{p6}
\end{figure}


\begin{thebibliography}{99}

\bibitem{venice}
T.~Toshito  [SuperKamiokande Collaboration],
arXiv:hep-ex/0105023. 

\bibitem{SK} 
S. Fukuda {\it et al.} [Super-Kamiokande Collaboration], 
Phys. Rev. lett. {\bf 86} (2001) 5651.  

\bibitem{SNO} 
Q. Ahmad {\it et al.} [SNO Collaboration], 
Phys. Rev. lett. {\bf 87} (2001) 071301.  

\bibitem{SNP}
For a recent review, see for example: 
M.~C.~Gonzalez-Garcia, M.~Maltoni, C.~Pena-Garay and J.~W.~F.~Valle,
Phys.\ Rev.\ D {\bf 63} (2001) 033005;\\
 G.~L.~Fogli, E.~Lisi, A.~Marrone, D.~Montanino and A.~Palazzo,
hep-ph/0104221.


\bibitem{BMW}V.~Barger, D.~Marfatia and K.~Whisnant, 
hep-ph/0106207; \\
%
J.~N.~Bahcall, M.~C.~Gonzalez-Garcia and C.~Pena-Garay,
hep-ph/0106258;\\
P.~I.~Krastev and A.~Y.~Smirnov, hep-ph/0108177
\bibitem{chooz}
M.~Apollonio {\it et al.}  [CHOOZ Collaboration],
Phys.\ Lett.\ B {\bf 466} (1999) 415.


\bibitem{nutev}
G.~P.~Zeller {\it et al.}  [NuTeV Collaboration],
hep-ex/0110059.

\bibitem{az3} Z.~Berezhiani and A.~Rossi, hep-ph/0111137.

\bibitem{AZ1}
Z.~Berezhiani and A.~Rossi,  Phys. \ Rev.\  D {\bf 51} (1995) 5229.

\bibitem{MSW}
S.~P.~Mikheev and A.~Yu.~Smirnov, Nuovo Cim.\ C {\bf 9} (1986) 17; \\
L. Wolfenstein, Phys.\ Rev.\ D {\bf 17} (1978) 2369. 
 
\bibitem{FC}
L.~Wolfenstein, ref. \cite{FC}; \\ 
J.~W.~F.~ Valle, Phys.\ Lett.\ B {\bf 199} (1987) 432; \\
E.~Roulet, Phys.\ Rev.\ D {\bf 44} (1991) 935; \\
M.~M.~Guzzo, A.~Masiero and S.~T.~Petcov,
Phys.\ Lett.\ B {\bf 260} (1991) 154.

\bibitem{BPW}
V.~Barger, R.~J.~Phillips and K.~Whisnant,
Phys.\ Rev.\ D {\bf 44} (1991) 1629.

\bibitem{AZ2}
Z.~G.~Berezhiani and A.~Rossi,
Proc. of the 5th Int. Workshop on `Neutrino Telescopes', p.   
123-135, ed. by M.~ Baldo Ceolin, Venice, Italy, 1993; hep-ph/9306278;  
Nucl.\ Phys.  \ Proc.\ Suppl.\  {\bf 35} (1994) 469.

\bibitem{analysis}
S.~Bergmann, M.~M.~Guzzo, P.~C.~de Holanda, P.~I.~Krastev and H.~Nunokawa,
Phys.\ Rev.\ D {\bf 62} (2000) 073001 and references 
therein.
\bibitem{venice1}
C.~McGrew [Super-Kamiokande Collaboration], Proc. of the  9th
Int. Workshop on `Neutrino Telescopes', vol.1, p.93, ed. by M.~Baldo Ceolin,
Venice, Italy, 2001.




\bibitem{donut}
M.~Nakamura  [DONUT Collaboration],
Nucl.\ Phys.\ Proc.\ Suppl.\  {\bf 77} (1999) 259. 
B.~Lundberg  [DONUT Collaboration],
Nucl.\ Phys.\ Proc.\ Suppl.\  {\bf 91} (2001) 233.



\bibitem{CHARM2}
P.~Vilain {\it et al.}  [CHARM-II Collaboration],
Phys.\ Lett.\ B {\bf 335} (1994) 246.

\bibitem{fornengo}
N.~Fornengo {\it et al.}, 
hep-ph/0108043.
\bibitem{FLM}
G.~L.~Fogli, E.~Lisi and A.~Marrone,
Phys.\ Rev.\ D {\bf 63} (2001) 053008; 
hep-ph/0105139.

\bibitem{BGP}
S.~Bergmann, Y.~Grossman and D.~M.~Pierce,
Phys.\ Rev.\ D {\bf 61} (2000) 053005; \\
S.~Bergmann {\it et al.}
Phys.\ Rev.\ D {\bf 62} (2000) 073001.






\bibitem{BKS}
J.~N.~Bahcall, M.~Kamionkowski and A.~Sirlin,
Phys.\ Rev.\ D {\bf 51} (1995) 6146.



\bibitem{BP00}
J.~Bahcall, M.~Pinsonneault and S.~Basu, astro-ph/0010346.


\bibitem{BX}
C. Arpesella {\it et al.} [BOREXINO Collaboration], Proposal of Borexino, 1991 
(unpublished); \\
G.~Ranucci {\it et al.}  [BOREXINO Collaboration],
Nucl.\ Phys.  \ Proc.\ Suppl.\  {\bf 91} (2001) 58.

\bibitem{taup95}
G.~Bellini [BOREXINO Collaboration], 
Nucl.\ Phys.  \ Proc.\ Suppl.\  {\bf 48} (1996) 363.

\bibitem{meroni}
E.~Meroni [BOREXINO Collaboration], 
Nucl.\ Phys. \ Proc.\ Suppl.\  {\bf 100} (2001) 42;
G.~Alimonti et al. [BOREXINO Collaboration], 
Astropart.\ Phys. {\bf 16} (2002) 205.

\bibitem{LENS}
R.~S.~Raghavan, 
Phys.\ Rev.\ Lett.  {\bf 78}, 3618 (1997).

\bibitem{GNHP}
M.~M.~Guzzo, H.~Nunokawa, P.~C.~de Holanda and O.~L.~Peres,
Phys.\ Rev.\ D {\bf 64} (2001) 097301.


\end{thebibliography}
\end{document}